\documentclass{egpubl}
\usepackage{pg2024}
\EGlocalpagenumber
\copyrightTextTitPag{\copyright\ 2024\ The Authors.\\
}
\def\@oddfoot{{\tiny\raisebox{\z@}[8pt][1pt]{\parbox[t]{20pc}{\sloppy}}}}
\def\@evenfoot{\mbox{}\hfill {\tiny\raisebox{\z@}[8pt][1pt]{\parbox[t]{20pc}{\sloppy}}}}
\def\ps@titlepage{\let\@mkboth\@gobbletwo
\def\@oddhead{\raisebox{\z@}[8pt][1pt]{\parbox{\textwidth}{\small\parbox[t]{.7\textwidth}{\sloppy\raggedright
Pacific Graphics 2024 /  R.\ Chen, T.\ Ritschel, and E.\ Whiting}
\hfill\parbox[t]{.3\textwidth}{\sloppy\raggedleft
\textit{Conference Paper}}}%
}%
}
\def\@oddfoot{{\tiny\raisebox{\z@}[8pt][1pt]{\parbox[t]{20pc}{\sloppy
\p@copyrightTextTitPag}}}\hfill}
}
\makeatother

\usepackage[T1]{fontenc}
\usepackage{dfadobe}  

\usepackage{cite}  
\BibtexOrBiblatex
\electronicVersion
\PrintedOrElectronic
\ifpdf \usepackage[pdftex]{graphicx} \pdfcompresslevel=9
\else \usepackage[dvips]{graphicx} \fi

\usepackage{egweblnk} 

\usepackage{booktabs}
\usepackage{amssymb}
\usepackage[dvipsnames, svgnames, x11names]{xcolor}  
\definecolor{my_green}{HTML}{0c8918} 
\definecolor{my_blue}{HTML}{177cb0}
\definecolor{my_orange}{HTML}{9b4400}
\definecolor{my_pruple}{HTML}{801dae}
\definecolor{my_red}{HTML}{be002f}
\definecolor{my_light_blue}{HTML}{00e5e6}
\newcommand{\etal}{\emph{et al.}}
\usepackage{multirow}
\usepackage{threeparttable}
\usepackage{amsmath}
\usepackage{cleveref}
\crefname{figure}{Fig.}{Figs.}
\Crefname{figure}{Figure}{Figures}
\crefname{table}{Tab.}{Tabs.}
\Crefname{table}{Table}{Tables}
\crefname{algorithm}{Alg.}{Algs.}
\Crefname{algorithm}{Algorithm}{Algorithms}
\crefname{equation}{Eq.}{Eqs.}
\crefname{section}{Sec.}{Secs.}
\usepackage{algpseudocode}
\usepackage{algorithm}
\makeatletter
\let\c@lofdepth\relax
\let\c@lotdepth\relax
\makeatother
\usepackage{graphicx}
\usepackage{subcaption} 
\title{PhysHand: A Hand Simulation Model with Physiological Geometry, Physical Deformation, and Accurate Contact Handling}

\author[M.Sun., D.Kou., \etal]
{\parbox{\textwidth}{\centering 
Mingyang Sun\thanks{Equal Contributions.}$^{,1}$\orcid{0009-0004-8217-6508},
Dongliang Kou$^{\dagger,1}$\orcid{0009-0009-5792-6895},
Ruisheng Yuan$^{1}$\orcid{0009-0005-3986-6287},
Dingkang Yang$^{1}$\orcid{0000-0003-1829-5671},
Peng Zhai$^{1}$\orcid{0000-0002-1374-7969},
Xiao Zhao$^{1}$\orcid{0000-0001-7928-3354},\\
Yang Jiang$^{1}$\orcid{0009-0006-5382-0303},
Xiong Li$^{4}$\orcid{0000-0001-5324-1404},
Jingchen Li$^{\ddagger,4}$\orcid{0000-0003-3196-7882},
and Lihua Zhang\thanks{Corresponding Authors.\\Email: jingchenli@tencent.com\\Email: lihuazhang@fudan.edu.cn
}$^{,1,2,3}$\orcid{0000-0003-0467-4347}
}\\
{\parbox{\textwidth}{\centering $^1$Academy for Engineering and Technology, Fudan University, Shanghai, China.\\
$^2$Engineering Research Center of AI and Robotics, Ministry of Education, Changchun, China.\\
$^3$Jilin Provincial Key Laboratory of Intelligence Science and Engineering, Changchun, China.\\
$^4$Tencent Robotics X Lab, Shenzhen, China.
       }
}
}

\begin{document}

\maketitle
\begin{abstract}
In virtual Hand-Object Interaction (HOI) scenarios, the authenticity of the hand's deformation is important to immersive experience, such as natural manipulation or tactile feedback.
Unrealistic deformation arises from simplified hand geometry, neglect of the different physics attributes of the hand, and penetration due to imprecise contact handling.
To address these problems, we propose \textbf{PhysHand}, a novel hand simulation model, which enhances the realism of deformation in HOI.
First, we construct a physiologically plausible geometry, a layered mesh with a ``skin-flesh-skeleton'' structure.
Second, to satisfy the distinct physics features of different soft tissues, a constraint-based dynamics framework is adopted with carefully designed layer-corresponding constraints to maintain flesh attached and skin smooth.
Finally, we employ an SDF-based method to eliminate the penetration caused by contacts and enhance its accuracy by introducing a novel multi-resolution querying strategy.
Extensive experiments have been conducted to demonstrate the outstanding performance of PhysHand in calculating deformations and handling contacts.
Compared to existing methods, our PhysHand: 
1) can compute both physiologically and physically plausible deformation;
2) significantly reduces the depth and count of penetration in HOI.



\end{abstract}   

\section{Introduction}
\label{sec1:intro}
In Hand-Object Interaction (HOI) scenarios, it is important to enhance the authenticity of the hand's deformation during contact, especially for tasks requiring fine manipulation in virtual environments\cite{sorli2021fineISMAR,delrieu2020fine} or deformation-dependent tactile feedback\cite{pham2017forceest, roke2012tactile1, zhang2021tactile2}.
Current research either overlooks deformations\cite{ott2010twohand,jacobs2012god,wang2024deepsimho}, simplifies the  geometry and dynamics\cite{verschoor2018softVR18,sorli2021fineISMAR,garre2011haptic}, or behaves poorly in handling contacts\cite{liu2023contactgen,jacobs2011softVR11,hirota2016interactionVR16}.
Although they have demonstrated potential applications in HOI, they struggle to meet the requirements of faithfully reproducing the deformation as in the real world.
Creating a virtual model that accurately approaches a real hand is indeed a complex task, involving challenges from geometry, dynamics, and contact, which will be sequentially discussed in the following.

\noindent \textbf{Geometry}. 
The human hand possesses a layered structure comprising bones, muscles, fat, and skin. 
Some works\cite{ott2010twohand,jacobs2012god} only consider the surface mesh of the skin, while others represent the skeleton with boxes\cite{garre2011haptic} or capsules\cite{sorli2021fineISMAR,verschoor2018softVR18}.
They all overlook a crucial fact: a comprehensive geometric model is fundamental for achieving realistic simulation results.
Although the internal bones do not need to be visualized, their shapes exactly influence the deformation of the soft tissues.

\noindent \textbf{Dynamics}. 
The motion of articulated skeletons can be easily computed through forward kinematics\cite{ott2010twohand} and rigid dynamics\cite{baraff1997rigid}, but the skin and flesh exhibit distinct non-rigid behaviors.
Some methods\cite{ott2010twohand,jacobs2012god,wang2024deepsimho} adopt skinning weights\cite{james2005skinning} to achieve coarse skin deformations, while others\cite{garre2011haptic,verschoor2018softVR18,sorli2021fineISMAR,jacobs2011softVR11} employ the Finite Element Method (FEM) as a dynamics framework.
However, to reduce computational costs, the FEM-based research treats the skin and flesh as identical soft bodies and employs a linear elastic model, resulting in excessively smooth deformations.

\noindent \textbf{Contact}. 
In simulation, contact handling (or collision detection) is a crucial component. It is used to calculate physical interactions between objects, striving to minimize or eliminate penetrations.
Rich contacts always imply overlap between different objects, posing challenges for accurate and fast collision detection and correction.
Mesh-based methods\cite{zhang2000meshcollision2,curtis2008meshcollison3,harmon2008meshCollision1} are poor in numerical stability due to the discontinuity in contact geometry\cite{werling2021collison4} and penetration\cite{todorov2012mujoco}.
Another option is querying the Signed Distance Function (SDF) to directly calculate the vector containing penetration depth and direction, which provides robustness for interaction.
The key problem is to find the point representing a face's contact status. Some works consider increasing the number of query points\cite{fuhrmann2003vertex-sdf,sampleSDF} or solving an optimization model\cite{macklin2020optsdf}.
The former results in a plethora of ineffective queries, while the latter is sensitive to initial values and prone to converge to local optima.

\begin{figure}[tb]
  \centering
\includegraphics[width=0.48\textwidth]{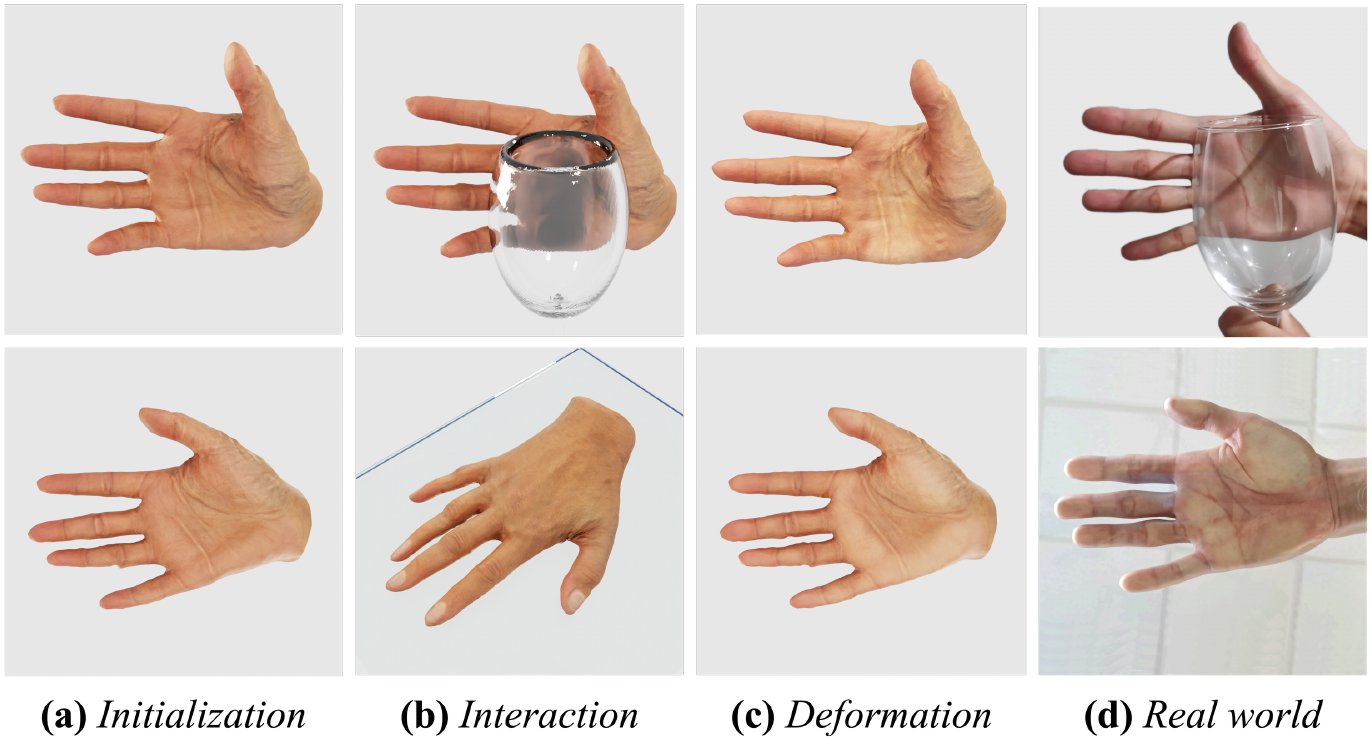}
  \caption{From left to right, we sequentially show the \textbf{initialization} without contact, the \textbf{interaction} between the hand and the object, the \textbf{deformation} of the hand without the object rendered, and the same contact and deformation in the \textbf{real world}. 
Our proposed \textbf{PhysHand} is capable of computing deformations of the hand that are almost consistent with the real world, thanks to our physiological geometry, physical deformation, and accurate contact handling.
We believe that Physhand will benefit downstream tasks such as immersive interaction, robust grasp, realistic tactility feedback, \textit{etc.}
}
\label{fig:teaser}
\end{figure}

The abovementioned problems can be summarized as inconsiderate geometry, simplified dynamics, and inaccurate contact handling.
To enhance the simulation quality of HOI, Our goal is to develop a hand simulation model to achieve realistic deformations in HOI (\cref{fig:teaser}).
\textbf{Firstly}, a hand object with physiological geometry is the fundamental element. NIMBLE\cite{li2022nimble} proposes a parametric function that fits data from MANO\cite{romero2017mano}, Magnetic Resonance Images (MRI), and texture images, resulting in a more comprehensive structure.
Based on NIMBLE, we additionally create a volumetric mesh as the flesh layer through Delaunay triangulation\cite{de2008delaunay}.
Consequently, we obtain a layered hand geometry that fully encompasses the skin, flesh, and skeleton, which adheres to physiology and lays the foundation for the subsequent simulation.
\textbf{Secondly}, a flexible dynamics framework is necessary to facilitate the design of the hand's physical characteristics.
Recently, a novel dynamics framework called eXtended Position Based Dynamics (XPBD)\cite{muller2007pbd,pbdEG,macklin2016xpbd} has been proposed, which allows users to establish constraints between vertices of the simulated object and solve them with the Gauss-Seidel method.
Constraints in XPBD are flexible, allowing for the establishment of any constraint between any vertices. Therefore, we adopt XPBD as the simulation backbone and design corresponding constraints for flesh and skin, ensuring physical plausibility.
\textbf{Finally}, accurate collision detection is a prerequisite for generating deformations. It is the most critical component in ensuring that contacts are handled correctly. 
Mesh-based representation is discrete and performs poorly when handling detailed contacts, whereas SDF is continuous and offers infinite resolution and is adopted by us. 
The colliding object is represented with an SDF, while the hand is modeled as a mesh. 
The target is leveraging SDF to identify points on the face of the mesh that accurately describe the contact state. 
To achieve this, we propose a novel multi-resolution querying strategy by iteratively querying on the face, significantly enhancing the accuracy of SDF-based contact handling. 
Additionally, this strategy naturally integrates with our dynamics framework by acting as a hard constraint, which is solved at once rather than iteratively\cite{macklin2020optsdf}.

Through our qualitative and quantitative analyses, as well as the comparative experiments, we demonstrate that our proposed PhysHand has the capability to calculate authentic deformations of the hand with both physiological and physical plausibility. Our contact handling strategy distinctly surpasses existing methods in terms of penetration depth and count. To summarize, the core competencies of PhysHand are supported by the following:
\begin{itemize}
\item A layered hand geometry containing the skin, flesh, and skeleton.

\item Layer-corresponding physics formulas within a constraint-based dynamics framework.

\item A multi-resolution querying strategy for accurate SDF-based contact handling.

\end{itemize}

\section{Related Works}
\label{sec2:rw}

We primarily review studies concerning geometric models, dynamics, and contact handling related to hands and conduct a corresponding comparison in \cref{tab:compare}.

\begin{table*}[tb] 
\caption{Comparison between PhysHand and other works. 
For the hand model, only PhysHand adopts a complete geometric structure comprising the skin, flesh, and skeleton, while others either lack some of elements\cite{ott2010twohand,jacobs2012god,wang2024deepsimho,jacobs2011softVR11,hirota2016interactionVR16} or replace them with simplified geometries\cite{garre2011haptic,verschoor2018softVR18,sorli2021fineISMAR}. 
For deformations, most studies employ skinning weights\cite{ott2010twohand,jacobs2012god,wang2024deepsimho} or solve linear elastic models\cite{garre2011haptic,jacobs2011softVR11,hirota2016interactionVR16,verschoor2018softVR18,sorli2021fineISMAR,murai2017dynamic_add}, while PhysHand create corresponding physics constraints for the flesh and skin.
Regarding contact handling, the mesh-based method lacks numerical stability, while Akihiko \etal\cite{murai2017dynamic_add} don't consider this aspect.
In contrast to existing SDF-based approaches\cite{fuhrmann2003vertex-sdf,sampleSDF,macklin2020optsdf}, PhysHand introduces a multi-resolution querying strategy that makes it more precise.
}
\begin{center}
\label{tab:compare}
\begin{threeparttable}
\centering
    {
    \begin{tabular}{ccccccccc}
    \toprule
        \multirow{2}{*}{Research} & \multicolumn{3}{c}{Hand Layer} & \multirow{2}{*}{Deformation}  & \multirow{2}{*}{Contact Handling} \\ 
        \cline{2-4}
        & Skin & Flesh & Skeleton  & ~ & ~  \\ 
        \midrule
        Ott \emph{et al.}\cite{ott2010twohand} & \checkmark & - & - & Skinning weights & Mesh \\ 
        Jacobs \emph{et al.}\cite{jacobs2012god} & \checkmark & - & - & Skinning weights & Mesh \\ 
        Wang \etal\cite{wang2024deepsimho} & \checkmark & - & - & Skinning weights & SDF \\
        Jacobs \emph{et al.}\cite{jacobs2011softVR11} & \checkmark & \checkmark& - & Linear elasticity & Mesh \\ 
        Hirota \emph{et al.}\cite{hirota2016interactionVR16} & \checkmark & \checkmark & - & Linear elasticity & Mesh \\ 
        Garre \emph{et al.}\cite{garre2011haptic} & \checkmark & \checkmark & Box & Linear elasticity & Mesh  \\ 
        Verschoor \emph{et al.}\cite{verschoor2018softVR18}& \checkmark & \checkmark  & Capsule  & Linear elasticity & SDF \\ 
        Sorli \emph{et al.}\cite{sorli2021fineISMAR} & \checkmark & \checkmark  & Capsule  & Linear elasticity & SDF \\
        Akihiko \etal\cite{murai2017dynamic_add} & \checkmark & \checkmark  & Mesh  & Linear elasticity & - \\
        PhysHand (ours) & \checkmark & \checkmark & Detailed mesh & Layer-corresponding constraints & SDF with multi-resolution querying\\
        \bottomrule
    \end{tabular}
    }
\end{threeparttable}
\end{center}
\end{table*}

\noindent \textbf{Geometric Models.}
Due to the complex structure of the hand, different approaches make a simplicity by considering only skin\cite{ott2010twohand,jacobs2012god,wang2024deepsimho}, fingers\cite{jacobs2011softVR11}, skin and flesh as a whole\cite{verschoor2018softVR18,hirota2016interactionVR16,garre2011haptic,sorli2021fineISMAR}, or a reduced skeleton composed of boxes\cite{garre2011haptic} or capsules\cite{verschoor2018softVR18}.
Various works\cite{qian2020html,li2022nimble,romero2017mano} have created hand models by collecting data from the real world and fitting them into a parametric function.
Among these models, NIMBLE\cite{li2022nimble} additionally fits data from MRI, enabling it to generate a complete skeleton.
Therefore, based on NIMBLE, we construct a physiologically layered hand model and make it suitable for simulation.

\noindent \textbf{Dynamics.}
The dynamic behavior can be calculated by geometric-based or physics-based deformation.
Geometric-based deformation\cite{wang2024deepsimho,qian2020html,romero2017mano} associates the positions of vertices with the joints of the skeleton through weights, known as skinning\cite{james2005skinning}. 
As the pose changes, vertices move with the joints according to the skinning weights. 
Geometric-based methods are convenient and fast, typically used for coarse deformation of the skin to match different poses.
Physics-based ways solve elasticity models to compute the displacement of vertices. 
Akihiko \etal\cite{murai2017dynamic_add} introduce spring–damper connections to calculate soft tissue dynamics, whose authenticity is dependent on the ability of the mass-spring model.
Although FEM incurs significant computational costs, it remains popular for simulating soft tissues\cite{comas2008fem1,kim2017fem2}. 
To reduce computational overhead, Garre \etal\cite{garre2011haptic} and Hirota \etal\cite{hirota2016interactionVR16} adopt linear elastic models. To enhance nonlinear behavior, Verschoor \etal\cite{verschoor2018softVR18} add a quickly growing energy term when the linear elastic energy density function reaches a specific threshold.
XPBD\cite{macklin2016xpbd,pbdEG,muller2007pbd} is a novel simulation framework that does not directly compute forces but solves constraints between vertices of simulated objects. As the iteration progresses, vertices' positions are updated directly to gradually satisfy the constraints.
Due to the flexibility in constraint design and the scalability of constraint combinations\cite{pbdEG}, XPBD has been utilized in simulations of human\cite{hi20203xpbdhuman,ye2022rcare}. 
Besides, the SDF-based contact handling method can be properly integrated into XPBD by forming the collision constraints.
Therefore, we adopt XPBD and design specific constraints for hands, ensuring excellent dynamic behaviors of the flesh attachment, and skin smoothness.

\noindent \textbf{Contact Handling.}
Typically, for two objects composed of triangular meshes, a mesh-based contact detection involves checking if any of their triangles intersect\cite{harmon2008meshCollision1,zhang2000meshcollision2,curtis2008meshcollison3}. 
However, it is lack of numerical stability because the meshes are discontinuous.
Another approach is to implicitly represent the surface of one of the objects using the zero-level-set of SDFs.
Querying the vertices of the other object can yield the signed distances and gradients, which are used to decide the contact states\cite{fuhrmann2003vertex-sdf}.
For a triangle of the mesh, the key issue is to find a point with the lowest signed distance.
Fuhrmann \emph{et al.}\cite{fuhrmann2003vertex-sdf} check and select the minimal signed distance from the vertices and the edges' center, but ignore the query within the triangle.
Macklin \emph{et al.}\cite{macklin2020optsdf} solve an optimization on the triangle, approaching the optimal point iteratively.
Nevertheless, due to the nonlinearity of SDFs, the results are susceptible to the initial point and converge towards local optima, leading to failed detection.
In contrast, we perform multi-resolution querying on triangles globally, achieving more precision.

\section{Proposed Method}
\label{sec3:method}
\begin{figure*}[htb]
  \centering
  \includegraphics[width=1\textwidth]{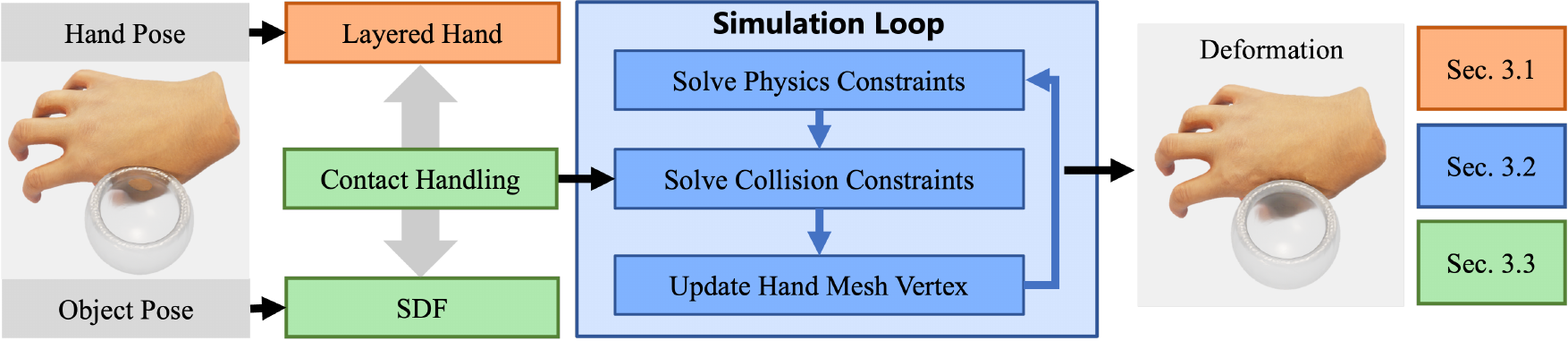}
  \caption{Schematic overview of the PhysHand. 
  The pose of hand and object can be obtained by manual setup or from a generative model\cite{liu2023contactgen}.
  The layered hand serves as a simulation entity, governed by the simulation loop.
  The simulation loop updates the vertices of the hand model by solving constraints, where the collision constraints are generated by the contact handling module when the hand comes into contact with the object.
}
  \label{fig:sec3-overview}
\end{figure*}

The main philosophy behind our method is to maximize the approximation of physical deformation in HOI in the real world.
In contrast to existing studies\cite{jacobs2011softVR11,hirota2016interactionVR16,verschoor2018softVR18,sorli2021fineISMAR,wang2024deepsimho}, we put special emphasis on modeling a detailed hand geometry, calculating non-rigid deformation of different part, and accurately handling contact caused by interaction.
To this end, in \cref{sec3.1:geo} we implement an extension of a parametric human hand model\cite{li2022nimble} to form a layered geometry that conforms to physiology. 
In \cref{sec3.2:dyn}, we design four constraints in a constraint-based dynamics framework\cite{macklin2016xpbd} for our hand geometry.
In \cref{sec3.3:con}, a novel contact handling method is proposed that enhances the performance of SDF-based collision detection\cite{fuhrmann2003vertex-sdf,macklin2020optsdf} by introducing a multi-resolution querying strategy.
\cref{fig:sec3-overview} depicts the schematic overview of our work. 

\subsection{Geometry: The Layered Hand Mesh}
\label{sec3.1:geo}
\begin{figure}[t]
\centering
    \begin{minipage}[t]{0.45\textwidth}
    \begin{subfigure}{0.22\textwidth}
        \centering
        \includegraphics[width=1\textwidth]{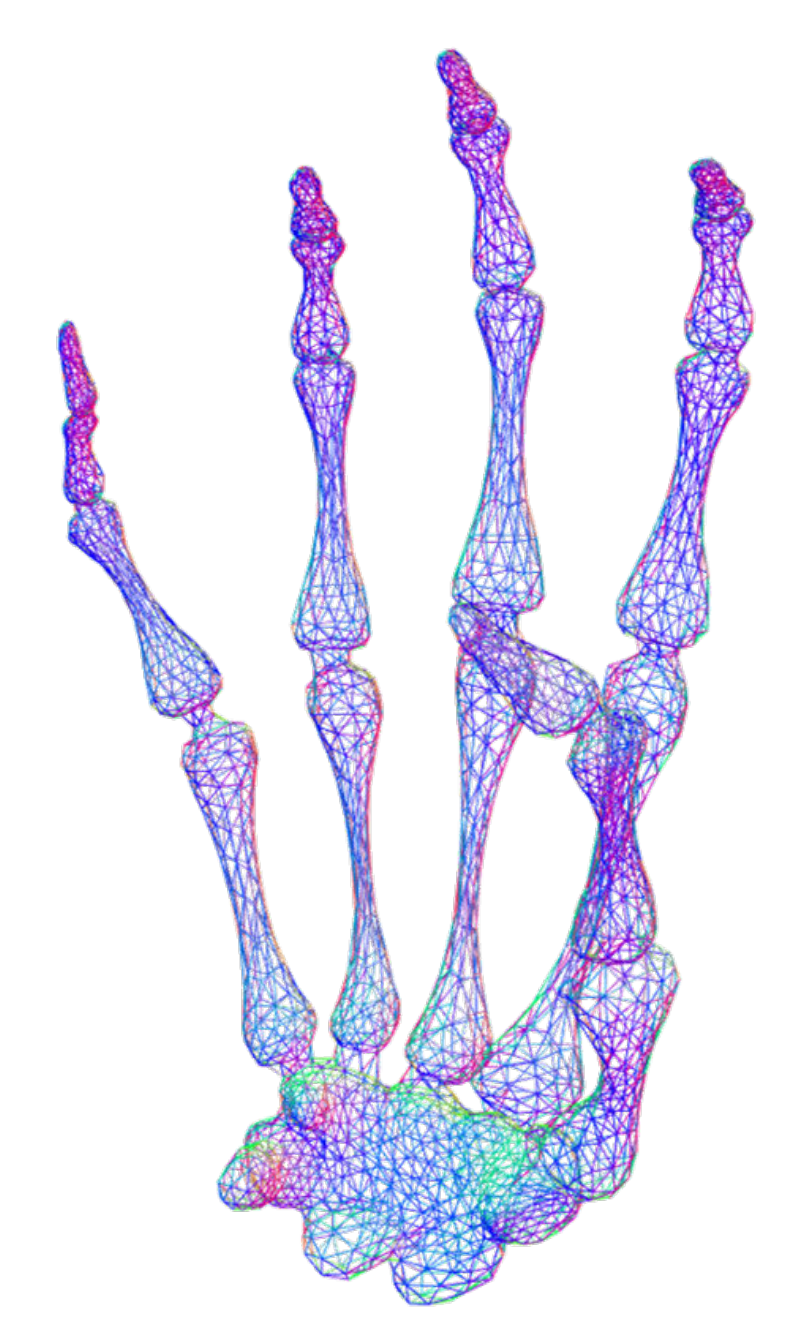}
        \caption{Skeleton}
        \label{fig:sec3.1-a}
    \end{subfigure}%
        \hfill
    \begin{subfigure}{0.22\textwidth}
        \centering
        \includegraphics[width=1\textwidth]{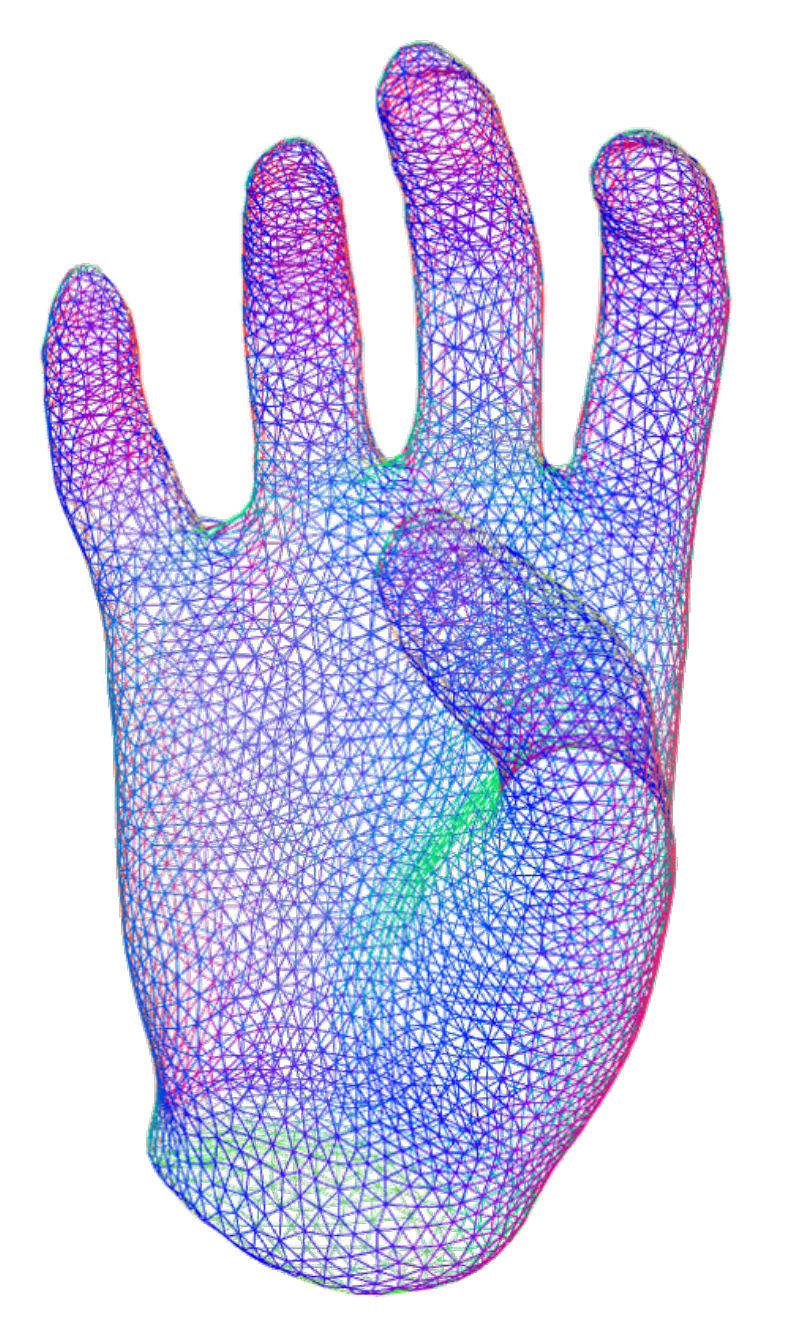}
        \caption{Skin}
        \label{fig:sec3.1-b}
    \end{subfigure}%
    \hfill
        \begin{subfigure}{0.22\textwidth}
        \centering
        \includegraphics[width=1\textwidth]{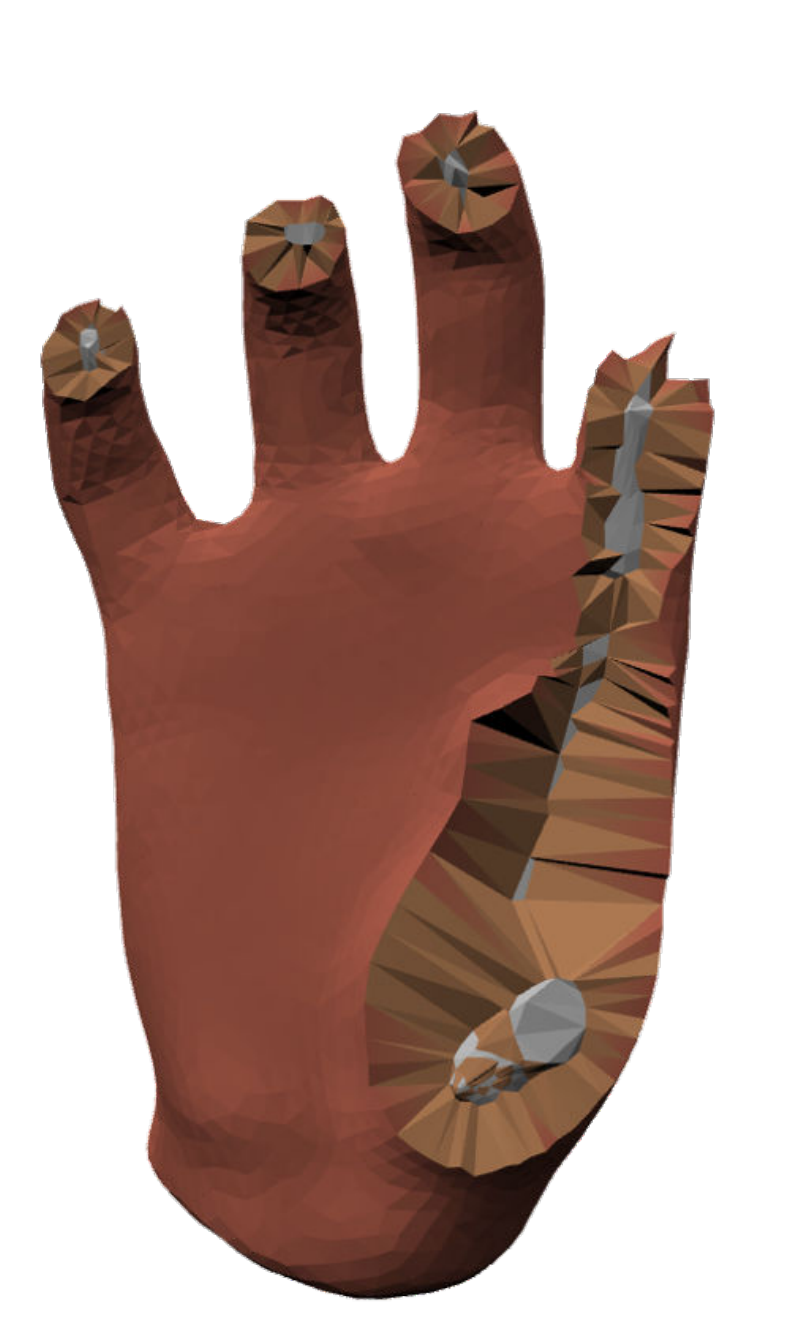}
        \caption{Flesh}
        \label{fig:sec3.1-c}
    \end{subfigure}%
    \end{minipage}

\caption{Illustration of the layered geometry.
    Both the skeleton and skin layers consist of triangular meshes, while the flesh layer is formed by tetrahedral meshes generated between them.
    \label{fig:sec3.1}
  }
\end{figure}

Let's begin by establishing this layered geometry. We follow the parametric function in NIMBLE\cite{li2022nimble}:
\begin{equation}
\label{eq1:nimble}
    \mathcal{G}(\theta,\beta)=\boldsymbol{LBS}(\mathcal W,J_{p},\theta,\bar T).
\end{equation}
We obtain the triangular mesh of the skin and skeleton from $\mathcal{G}$.
$\boldsymbol{LBS}(\cdot)$ denotes the Linear Blend Skinning (LBS) function; $\mathcal{W}$ is the skinning weight of $\boldsymbol {LBS}(\cdot)$; $J_p$ represents the joint locations; $\theta$ is the pose of the skeleton; $\beta$ is the Principal Component Analysis (PCA) coefficient vector of the shape space; and $\bar T$ is the template mesh. For more details, please refer to \cite{li2022nimble}.
We set the values of $\beta$ and $\theta$ to 0, indicating the template with the rest pose. 
Then, we apply Delaunay triangulation \cite{de2008delaunay,hang2015tetgen} between the skeleton layer and the skin layer to fill tetrahedrons as the flesh layer.
Despite the diversity of soft tissues between the skin and bones, such as muscles, fat, and others, we argue that modeling these tissues as a single type of non-rigid flesh is sufficient for calculating realistic deformations compared to existing works\cite{sorli2021fineISMAR,verschoor2018softVR18}. 
Such a layered geometry adheres to physiology well. 
Specifically, skin is used for contact handling, the flesh is the primary deformable component, and the skeleton provides movement and rigid support.
Additionally, we also adopt skinning weights of the skeletal vertices derived from NIMBLE\cite{li2022nimble}, which is later utilized for calculating rigid body motions of the skeleton layer.
Consequently, we obtain a physiologically plausible geometry, as depicted in the \cref{fig:sec3.1}.
The complete hand model comprises 8,397 vertices and 9,450 triangular faces with an average area of 16.758 $ mm^2 $.

\subsection{Dynamics: Layer-corresponding Physics Constraints}
\label{sec3.2:dyn}
We adopt XPBD\cite{macklin2016xpbd} as the simulation framework, which establishes constraints between vertices.
In each time step, the positions of the vertices are projected onto each constraint manifold along the constraint gradient. Additionally, each constraint has a corresponding compliance, with lower values indicating greater rigidity.
For clarity, we provide the implementation in our method, as shown in \cref{alg:xpbd}. $\Delta t$ is the time step size and $h$ is the substep size. $\mathrm{\mathbf{x}}$, $\mathrm{\mathbf{v}}$, and $\mathrm{\mathbf{M}}$ represent the position, velocity, and mass of the particles, respectively.
The external force acting on the particles is represented as $\mathrm{\mathbf{f}}_\mathrm{ext}$, which in this paper refers to gravity.
Constraint compliance is denoted by $\alpha$, representing the strength of the constraint, which is independent of each constraint to achieve diverse physical characteristics.
$\boldsymbol{C}(\mathbf{x})$ is the constraint applied to the position of the particles, which is iteratively satisfied through the projection.
In this paper, the constraints include basic constraints of the edge and tetrahedron, as well as constraints corresponding to the layers of flesh and skin.
Specifically, we first update the position considering only external forces (\cref{alg:xpbd}.\ref{algline:1-3}). 
Next, collision detection and correction (\cref{alg:xpbd}.\ref{algline:1-6}) are employed to mitigate penetration.
Furthermore, based on the updated position, we calculate displacements required to satisfy constraints (\cref{alg:xpbd}.\ref{algline:1-8} to \cref{alg:xpbd}.\ref{algline:1-10}) then continue to correct the position (\cref{alg:xpbd}.\ref{algline:1-12}). 
Finally, velocities are updated (\cref{alg:xpbd}.\ref{algline:1-15}) based on the final positions (\cref{alg:xpbd}.\ref{algline:1-14}).

\begin{algorithm}[t]
\caption{XPBD simulation for a single time step.}\label{alg:xpbd}
\begin{algorithmic}[1]
\State $h \gets \Delta t / \mathrm{numSubSteps}$ \Comment{Substep size} \label{algline:1-1}
\For{$ i=1,\dots,\mathrm{numSubSteps} $} \Comment{Iteration} \label{algline:1-2}
    \State initialize solution $ \mathrm{\mathbf{x}}\gets \mathrm{\mathbf{x}}^i + h\mathrm{\mathbf{v}}^i + h^2\mathrm{\mathbf{M}^{-1}}\mathrm{\mathbf{f}}_\mathrm{ext}(\mathrm{\mathbf{x}}^i) $ \label{algline:1-3}
    \State initialize multipliers $ \mathrm{\mathbf{\lambda}}\gets 0 $ \label{algline:1-4}
    \State $ \mathrm{\mathbf{\tilde{\alpha}}} \gets \frac{1}{h^2}\alpha $ \label{algline:1-5}
    \State collision detection and correction \Comment{Sec. \ref{sec3.3:con}} \label{algline:1-6}
    \ForAll{constraints} \label{algline:1-7}
        \State $ \mathrm{\mathbf{A}} \gets  \nabla\boldsymbol{C}(\mathrm{\mathbf{x}})\mathrm{\mathbf{M}}^{-1}\nabla\boldsymbol{C}^T(\mathrm{\mathbf{x}}) + \mathrm{\mathbf{\tilde{\alpha}}} $ \label{algline:1-8}
        \State $ \Delta \mathrm{\mathbf{\lambda}} \gets -\mathrm{\mathbf{A}}^{-1}(\boldsymbol{C}(\mathrm{\mathbf{x}})+\mathrm{\mathbf{\tilde{\alpha}}} 
        \mathrm{\mathbf{\lambda}}) $ \label{algline:1-9}
        \State $ \Delta \mathrm{\mathbf{x}} \gets \mathrm{\mathbf{M}}^{-1} \nabla\boldsymbol{C}^T(\mathrm{\mathbf{x}})\Delta \mathrm{\mathbf{\lambda}}$ \Comment{Projection} \label{algline:1-10}
        \State $ \mathrm{\mathbf{\lambda}} \gets \mathrm{\mathbf{\lambda}} + \Delta \mathrm{\mathbf{\lambda}} $  \label{algline:1-11}
        \State $ \mathrm{\mathbf{x}} \gets \mathrm{\mathbf{x}} + \Delta\mathrm{\mathbf{x}} $ \label{algline:1-12}
    \EndFor \label{algline:1-13}
    \State update positions $\mathrm{\mathbf{x}}^{i+1} \gets \mathrm{\mathbf{x}}$ \label{algline:1-14}
    \State update velocities $\mathrm{\mathbf{v}}^{i+1} \gets \frac{\mathrm{\mathbf{x}}^{i+1}-\mathrm{\mathbf{x}}^i}{h}$ \label{algline:1-15}
    
    \State $i \gets i+1$ \label{algline:1-16}
\EndFor \label{algline:1-17}
\end{algorithmic}
\end{algorithm}

For better readability, we define $\mathbf x_{i,j}=\mathbf x_i - \mathbf x_j$.
Given an edge with vertices of $\mathbf{x}_1$ and $\mathbf{x}_2$ and rest length $L$, the edge constraint is defined as:
\begin{equation}
\label{eq:cedge}
    \boldsymbol{C}^\mathrm{edge}(\mathbf{x}_1,\mathbf{x}_2)=\parallel \mathbf{x}_{1,2} \parallel - L,
\end{equation}
where $\parallel \cdot \parallel$ means Euclidean norm.

For a tetrahedron with vertices $\mathbf{x}_1$, $\mathbf{x}_2$, $\mathbf{x}_3$ and $\mathbf{x}_4$, and rest volume $V$, we define the tetrahedron constraint as:
\begin{equation}
\label{eq:ctet}
    \boldsymbol{C}^\mathrm{tet}(\mathbf{x}_1, \mathbf{x}_2, \mathbf{x}_3, \mathbf{x}_4)= \mathbf{x}_{2,1}\times \mathbf{x}_{3,1}\cdot \mathbf{x}_{4,1} - 6V.
\end{equation}

As the basic constraints to ensure deformations, we apply $\boldsymbol{C}^\mathrm{edge}$ to all edges and $\boldsymbol{C}^\mathrm{tet}$ to all tetrahedra.
In particular, we set the inverse mass $\mathbf{M}^{-1}$ of vertices in the skeleton layer to 0, which in XPBD\cite{macklin2016xpbd} means their positions are not affected by constraints, ensuring each bone remains rigid.
However, $\boldsymbol{C}^\mathrm{edge}$ and $\boldsymbol{C}^\mathrm{tet}$ alone are insufficient to capture the characteristics of the hand.
Therefore, we further propose the corresponding constraints for the flesh layer and skin layer.
\begin{figure}[tb]
\centering
    \begin{subfigure}{0.21\textwidth}
        \centering
        \includegraphics[width=\textwidth]{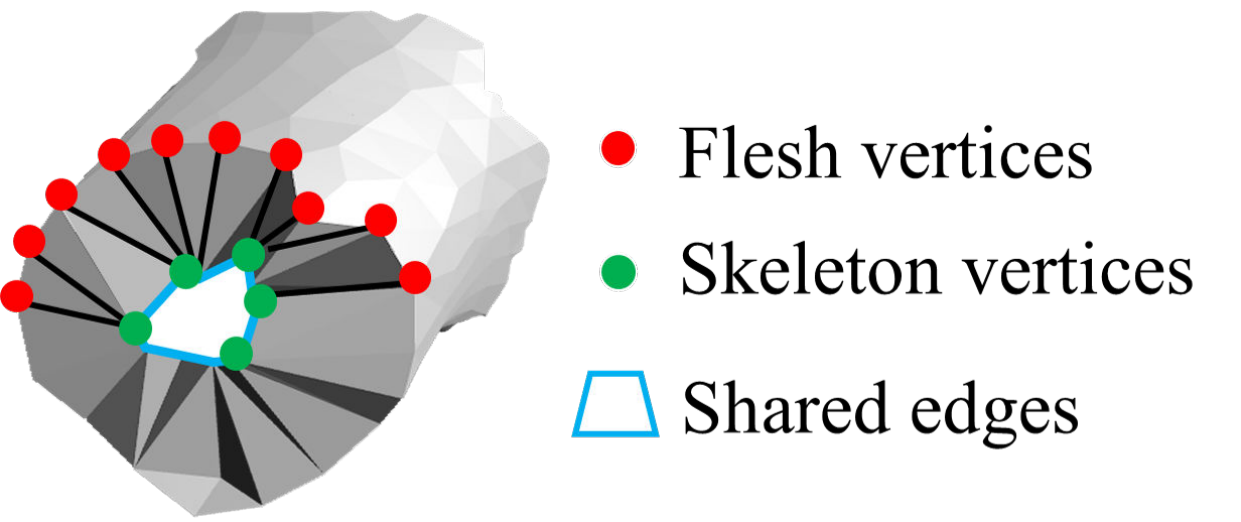}
        \caption{$\boldsymbol{C}^\mathrm{att}$}
        \label{fig:sec3.2-a-att}
    \end{subfigure}%
\hfill
    \begin{subfigure}{0.23\textwidth}
        \centering
        \includegraphics[width=\textwidth]{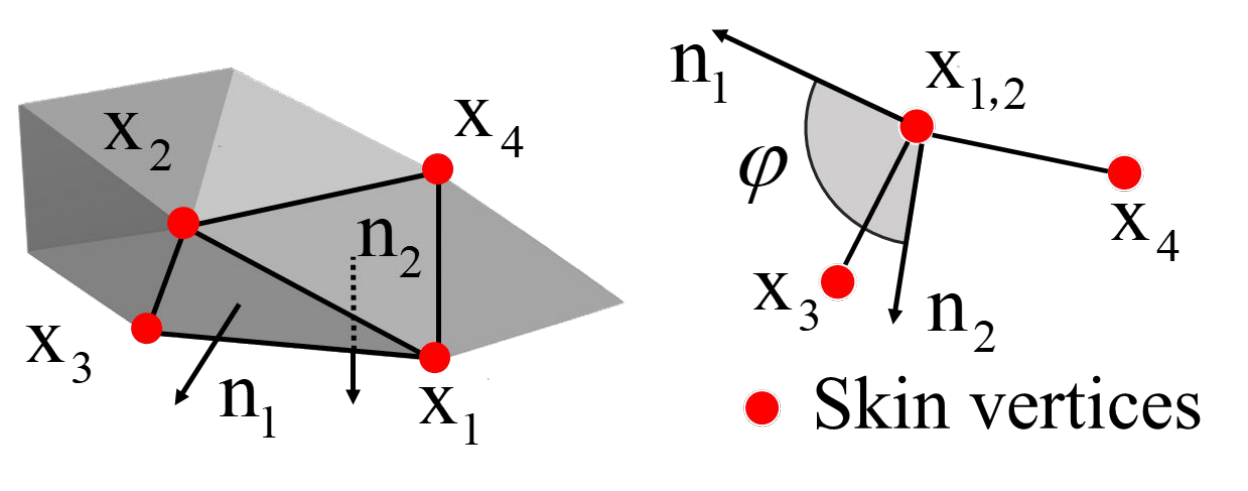}
        \caption{$\boldsymbol{C}^\mathrm{smo}$}
        \label{fig:sec3.2-b-smo}
    \end{subfigure}%
  	\caption{Illustration of physics constraints. 
    (a) Each vertex of the flesh layer establishes an edge constraint with the nearest skeleton vertex to prevent excessive deformation.
    (b) For the skin layer, we keep the dihedral angle of adjacent triangles through normal vectors to provide smoothness.}
   	\label{fig:sec3.2}
\end{figure}

\noindent \textbf{Flesh Attachment Constraint.}
For those vertices of bones and skin located within the same tetrahedron or along an edge, edge constraints and tetrahedron constraints make physical connections between the skeleton and skin, thereby preserving the shape of the flesh.
However, in practice, we have found that merely establishing tetrahedral constraints is insufficient, especially when the number of iterations is limited for real-time performance.
It results in the undesirable loosening of flesh.
Therefore, additional constraints need to be established.
As shown in \cref{fig:sec3.2-a-att}, for each flesh vertex $\mathbf{x}_\mathrm{flesh}$, we find its nearest skeleton vertex $\mathbf{x}_\mathrm{skeleton}$ and form an edge constraint between them.
Specifically, we define the flesh attachment constraint as:
\begin{align}
    \label{eq:catt}
    \boldsymbol{C}^\mathrm{att}(\mathbf{x}_\mathrm{flesh}) = \boldsymbol {C}^\mathrm{edge}(\mathbf{x}_\mathrm{flesh},\mathbf{x}_\mathrm{skeleton}),\\
    \textbf{x}_\mathrm{skeleton} = \arg\min_{\textbf{x}_i \in \boldsymbol{S}_\mathrm{skeleton}} ||\textbf{x}_{\mathrm{flesh},i}||,
\end{align}
where $\boldsymbol{S}_\mathrm{skeleton}$ is the set of vertices belong to the skeleton.

\noindent \textbf{Skin Smoothness Constraint.}
The deformation of flesh can cause the skin to become uneven, which goes beyond the natural wrinkles of the skin.
To alleviate distortion on the skin, we advocate maintaining the initial dihedral angle between interconnected triangles, referred to as the skin smoothness constraint (see \cref{fig:sec3.2-b-smo}):
\begin{equation}
\label{eq:csmo}
    \boldsymbol{C}^\mathrm{smo}(\mathbf{x}_1, \mathbf{x}_2, \mathbf{x}_3, \mathbf{x}_4)= \operatorname{\arccos}{(\mathbf{n}_1\cdot \mathbf{n}}_2)-\varphi,
\end{equation}
where $\mathbf x_1$, $\mathbf x_2$, $\mathbf x_3$, and $\mathbf x_4$ are vertices that belong to two triangles with a shared edge, $\mathrm{\mathbf{n}}_1$ and $\mathrm{\mathbf{n}}_2$ are triangle normals and $\varphi$ is the initial dihedral angle. We apply $\boldsymbol{C}^\mathrm{smo}$ to the skin.

\begin{figure*}[htbp]
\centering
\begin{minipage}[t]{\textwidth}
    \begin{subfigure}{0.245\textwidth}
        \centering
        \includegraphics[width=0.98\textwidth]{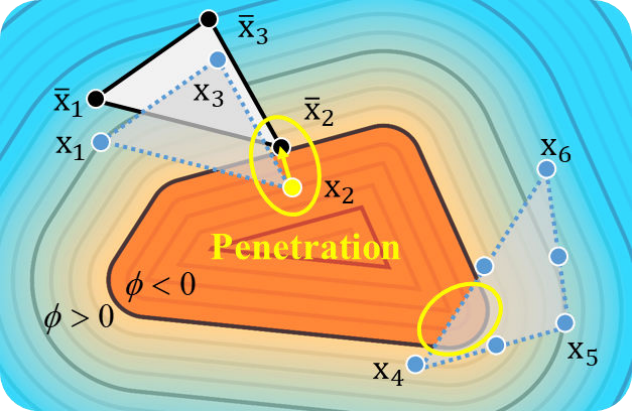}
        \subcaption{\textit{vertex-SDF}\cite{fuhrmann2003vertex-sdf}}
        \label{fig:sec3.3-a}
    \end{subfigure}%
\hfill
    \begin{subfigure}{0.245\textwidth}
        \centering
        \includegraphics[width=1.0\textwidth]{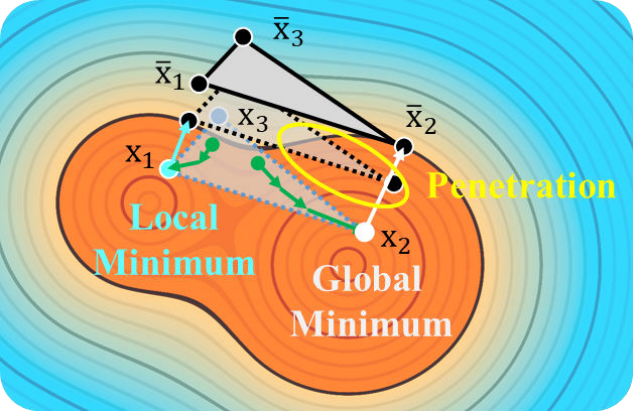}
        \subcaption{\textit{opt-SDF}\cite{macklin2020optsdf}}
        \label{fig:sec3.3-b}
    \end{subfigure}%
\hfill
    \begin{subfigure}{0.455\textwidth}
        \centering
        \includegraphics[width=0.96\textwidth]{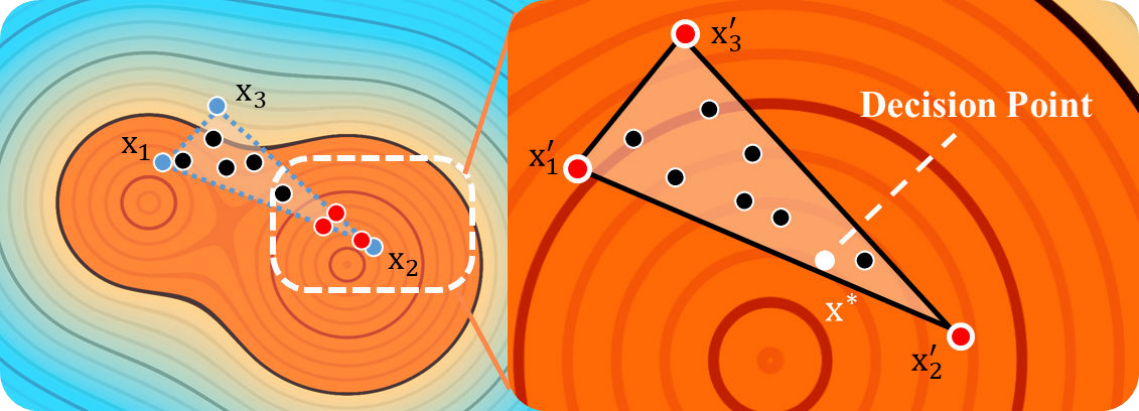}
        \subcaption{\textit{multi-res-SDF} (ours)}
        \label{fig:sec3.3-c}
    \end{subfigure}%
\end{minipage}
\caption{The comparison of contact handling methods based on SDF. 
    We define several triangles as $T_1=\{\mathbf{x}_1,\mathbf{x}_2,\mathbf{x}_3\}$, $T_2=\{\mathbf{x}_4,\mathbf{x}_5,\mathbf{x}_6\}$, $\Bar{T}_1=\{\mathbf{\Bar{x}}_1,\mathbf{\Bar{x}}_2,\mathbf{\Bar{x}}_3\}$, and $T_1^\prime=\{\mathbf{x}^\prime_1,\mathbf{x}^\prime_2,\mathbf{x}^\prime_3\}$, which are common to all subgraphs.
    (a) The black boundary between the \textcolor{DarkOrange}{orange} and \textcolor{ProcessBlue}{blue} regions represents the zero-level-set of the SDF.
    The triangles that experience penetration are transparent, such as $T_1$ and $T_2$. 
    According to the signed distance and gradient at $\mathbf{x}_2$, $T_1$ can be corrected to $\Bar{T}_1$.
    For $T_2$, the signed distances of its vertices and centers of edges fail to precisely depict its penetration state, constituting a drawback inherent to the \textit{vertex-SDF}.
    (b) \textit{opt-SDF} iteratively solves for an optimal point (\textcolor{ForestGreen}{green} points and arrows) with the lowest signed distance. 
    However, SDF is usually nonlinear, which makes \textit{opt-SDF} sensitive to initial guesses and susceptible to converging towards local optima (\textcolor{my_light_blue}{bright blue} point), leading to an insufficient correction (\textcolor{my_light_blue}{bright blue} arrow).
    Instead, the global optimum (white point) can help eliminate penetration. 
    (c) We perform global sampling on the triangle (left), then select three points (\textcolor{Red}{red} points) with the lowest signed distance to form a new triangle $T^\prime_1$ (right), iteratively increasing the resolution.  
    After the final sampling iteration, we select the point with the smallest signed distance as the decision point $\mathbf{x}^\ast$, 
    which is more likely to approach the global optimum.
}
  \label{fig:sec3.3-contacts}
\end{figure*}

\subsection{Contact Handling: The Multi-resolution Querying}
\label{sec3.3:con}
We employ an SDF-based approach to handle contact and propose a multi-resolution querying strategy to improve accuracy.

\noindent \textbf{SDF Representation.}
The zero-level-set of an SDF can implicitly describe the surface of an object:
\begin{equation}
\label{eq:sdf}
    \mathcal{S} = \{ \phi(\mathrm{\mathbf{x}}) \} = 0,
\end{equation}
where $\mathcal{S}$, $\phi$, and $\mathrm{\mathbf{x}}$ represent respectively the surface, SDF, and query point.
Most shapes can be represented as SDFs using analytical expressions\cite{sdfshader}.
For complex shapes, an available option is to train a neural SDF to fit the surface\cite{park2019deepsdf}.

\noindent \textbf{Multi-resolution Querying.}
Given a query point $\mathrm{\mathbf{x}}$, we define the collision constraint as:
\begin{equation}
\label{eq:coll_constraint}
    \boldsymbol{C}^\mathrm{coll}(\mathrm{\mathbf{x}}) = \phi(\mathrm{\mathbf{x}}).
\end{equation}
When $\boldsymbol{C}^\mathrm{coll}(\mathbf{x}) \le 0 $, penetration occurs, then $\mathrm{\mathbf{x}}$ needs to be projected to $\mathrm{\mathbf{x}}^\prime$ that satisfies the collision constraint:
\begin{equation}
\label{eq:coll_project}
    \mathrm{\mathbf{x}}^\prime = \mathrm{\mathbf{x}} - \frac{\nabla\phi(\mathrm{\mathbf{x}})}{||\nabla\phi(\mathrm{\mathbf{x}})||}\phi(\mathrm{\mathbf{x}}),
\end{equation}
where $\phi(\mathrm{\mathbf{x}})$, $\nabla\phi(\mathrm{\mathbf{x}})$, and $\mathbf{x}^\prime$ represent the SDF, the gradient at the query point x, and the corrected position, respectively.

For a triangle, the signed distance of its vertices cannot accurately describe whether it has penetrated or not. Additional queries of its edges' centers make a small difference\cite{fuhrmann2003vertex-sdf}, as shown in \cref{fig:sec3.3-a}.
To address this issue, we need to search for the point on the triangle with the lowest signed distance value. As long as the signed distance of this point is positive, the entire triangle is penetration-free. 
We call this point the \textit{decision point}.
Macklin \etal\cite{macklin2020optsdf} proposed an optimization model to iteratively approximate the decision point on a triangle (see \cref{fig:sec3.3-b}).
However, due to the nonlinearity of the SDF, this method is sensitive to the initial points and prone to converge to a local minimum.
Thus, we decide to replace optimization with a globally iterative multi-resolution sampling scheme (see \cref{fig:sec3.3-c}).
Given a triangle, we first sample $N$ points within its interior, including its vertices and edges.
Subsequently, we select the three points with the lowest signed distances to construct a new triangle and repeat the sampling and querying process. 
As the iteration progresses, the resolution grows quadratically. Thus, we call this strategy \textit{multi-resolution querying}.
Finally, we select the point with the lowest signed distance as the \textit{decision point} $\mathrm{\mathbf{x}}^\ast$. For decision points whose signed distance is less than zero, we will modify all vertices of the corresponding triangle according to \cref{eq:coll_project}.
The complete algorithm is outlined in \cref{alg:contact}.

\section{Experiments}
\label{sec4:exp}
The following questions will be mainly verified through both quantitative and qualitative experiments:
\begin{itemize}

\item How do layer-corresponding constraints affect the physical characteristics of the hand?
\item Compared with existing methods, can our contact handling strategy demonstrate higher accuracy?
\item How does PhysHand contribute to the existing research on HOI?
\item Can PhysHand replicate contact deformations similar to those in the real world?
\end{itemize}

\begin{figure}[tb]
\centering
    \begin{subfigure}{0.14\textwidth}
        \centering
        \includegraphics[width=0.85\textwidth]{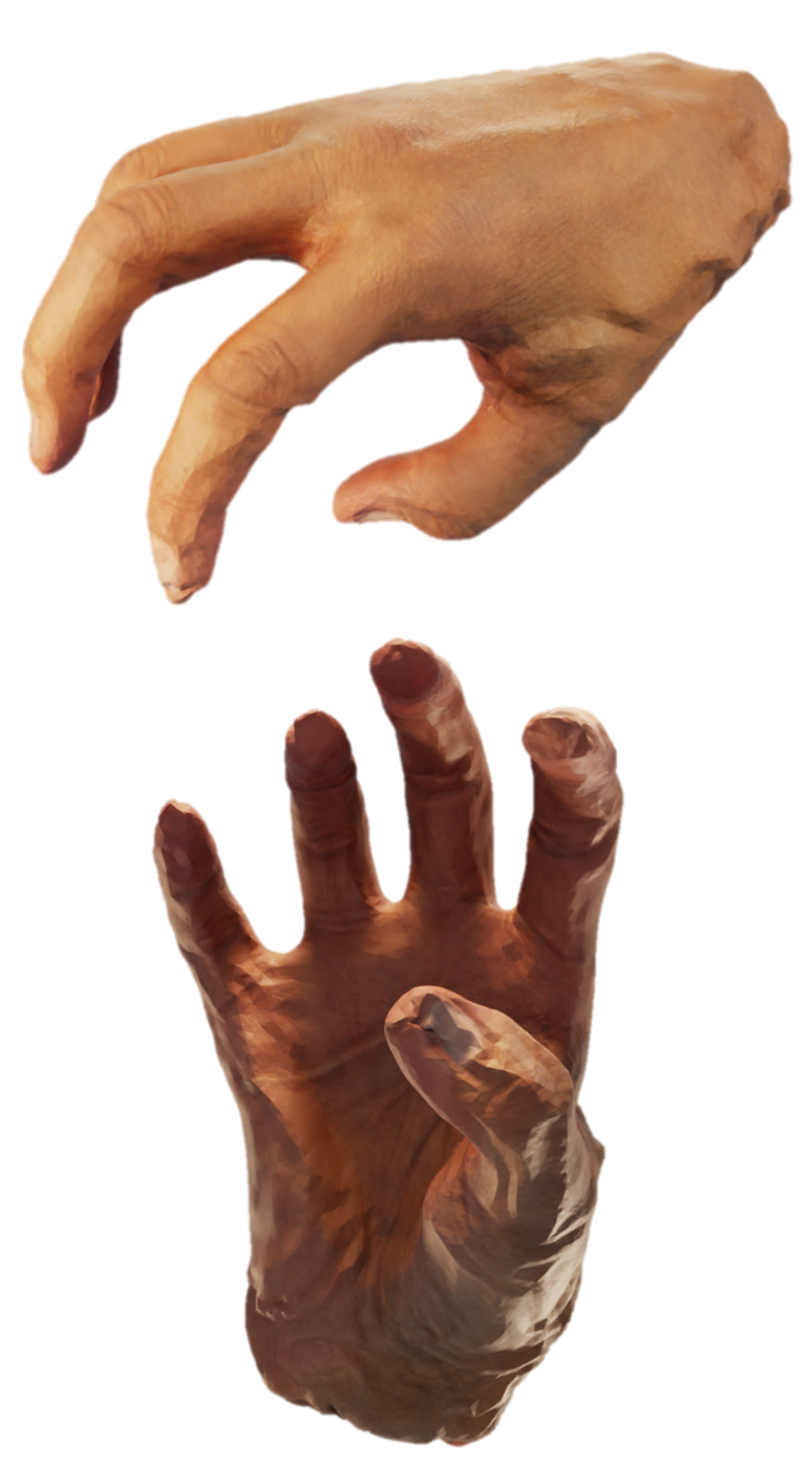}
        \caption{+ $\boldsymbol{C}^\mathrm{edge}$ and $\boldsymbol{C}^\mathrm{tet}$}
        \label{fig:sec4.1-a}
    \end{subfigure}%
 \hfill
    \begin{subfigure}{0.12\textwidth}
        \centering
        \includegraphics[width=0.99\textwidth]{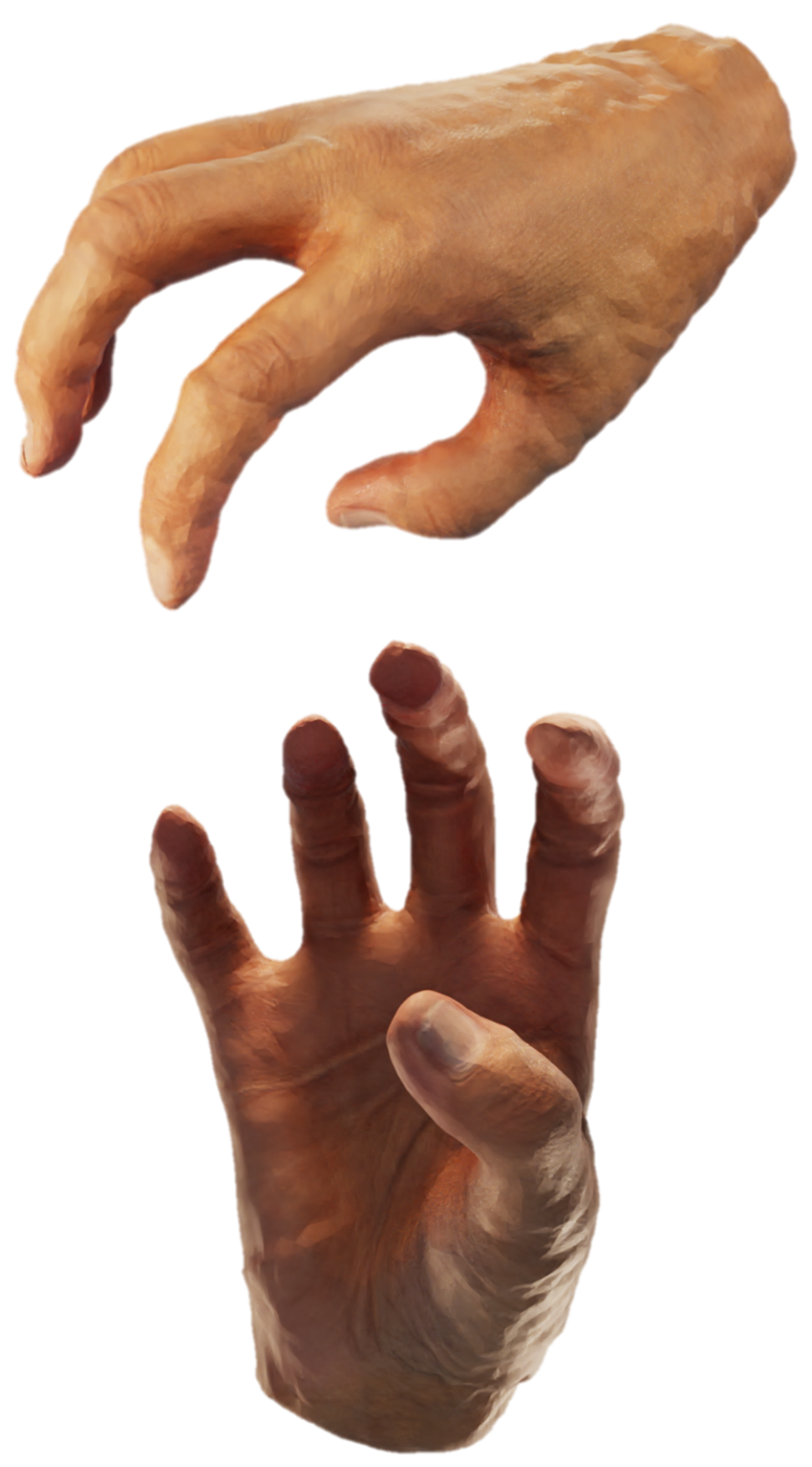}
        \caption{+ $\boldsymbol{C}^\mathrm{att}$}
        \label{fig:sec4.1-b}
    \end{subfigure}%
 \hfill
    \begin{subfigure}{0.12\textwidth}
        \centering
        \includegraphics[width=0.99\textwidth]{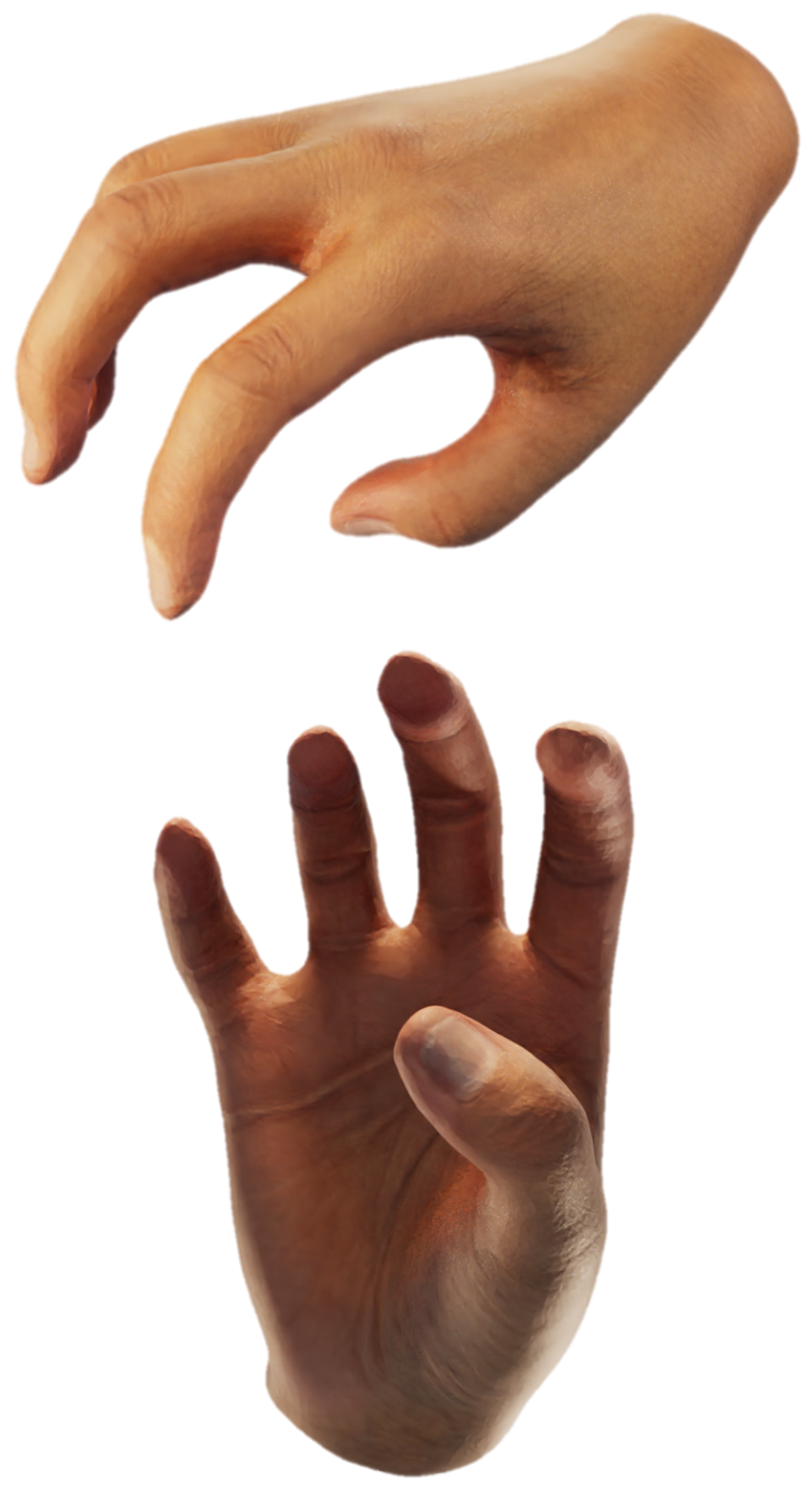}
        \caption{+ $\boldsymbol{C}^\mathrm{smo}$}
        \label{fig:sec4.1-c}
    \end{subfigure}%
        \caption{Improvements in deformation from layer-corresponding constraints. 
        We gradually add constraints.
        (a) The hand with $\boldsymbol{C}^\mathrm{edge}$ and $\boldsymbol{C}^\mathrm{tet}$. 
        (b) The activation of $\boldsymbol{C}^\mathrm{att}$. 
        (c) Enabling $\boldsymbol{C}^\mathrm{smo}$.
        }
   	\label{fig:sec4.1-constraints}
\end{figure}

\begin{algorithm}[t]
\caption{Contact handling with multi-resolution querying.}\label{alg:contact}
\begin{algorithmic}[1]
    \ForAll{triangles of the skin} 
        \State $T=\{  \mathbf{x}_1,\mathbf{x}_2,\mathbf{x}_3\}$
        \For {i = 1,$\dots$, sampling times}
            \State query the sampled points $\mathcal X$ on triangle $T$ 
            \State obtain the triangle $T^\prime = \{ \mathbf{x}_1^\prime,\mathbf{x}_2^\prime,\mathbf{x}_3^\prime\} $ with the lowest three signed distances.
            \State $T \gets T^\prime$
        \EndFor
        \State obtain the \textit{decision point} $\mathrm{\mathbf{x}}^\ast \gets \mathop{\operatorname{argmin}}\limits_{\mathrm{\mathbf{x}}} \phi(\mathrm{\mathcal{X}})$
        \If{$\phi(\mathrm{\mathbf{x}}^\ast)\leq 0$}
            \ForAll{$\mathbf x \in T=\{\mathbf{x}_1,\mathbf{x}_2,\mathbf{x}_3\}$}
                \State $\mathrm{\mathbf{x}} \gets \mathrm{\mathbf{x}} -\frac{ \nabla\phi(\mathrm{\mathbf{x}}^\ast)}{|| \nabla\phi(\mathrm{\mathbf{x}}^\ast)||}\phi(\mathrm{\mathbf{x}}^\ast)$ \Comment{\cref{eq:coll_project}}
            \EndFor
        \EndIf
    \EndFor
\end{algorithmic}
\end{algorithm}

\subsection{Constraints}
\label{sec4.1:exp-constraint}
We fix the skeleton with a rest pose and set the gravity acceleration to 9.8 $m/s^2$. 
The constraint formulas and their scope (sets of vertices) have been explained in \cref{sec3.2:dyn}. 
Initially, we activate only $\boldsymbol{C}^\mathrm{edge}$ and $\boldsymbol{C}^\mathrm{tet}$, followed by $\boldsymbol{C}^\mathrm{att}$, and finally enable $\boldsymbol{C}^\mathrm{smo}$. 
In \cref{fig:sec4.1-a}, the hand fails to maintain its shape due to the excessive deformation of the flesh.
\cref{fig:sec4.1-b} shows that the flesh layer becomes firm and full because we establish $\boldsymbol{C}^\mathrm{att}$ between the skin and the skeleton.
Due to the modification of dihedral angles generated from $\boldsymbol{C}^\mathrm{smo}$, the skin becomes smoother (see \cref{fig:sec4.1-c}).
Consequently, these constraints collectively ensure physical plausibility.

\begin{figure*}[htbp]
\begin{minipage}[t]{\textwidth}
\centering
    \begin{subfigure}{0.22\textwidth}
        \centering
        \includegraphics[width=1\textwidth]{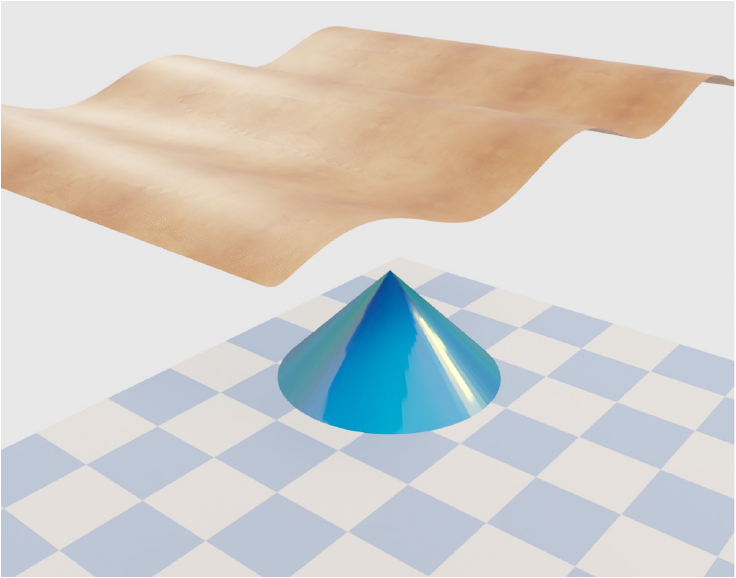}
        \caption{Initialization}
        \label{fig:sec4.2-vis-a}
    \end{subfigure}%
\hfill
    \begin{subfigure}{0.22\textwidth}
        \centering
        \includegraphics[width=1\textwidth]{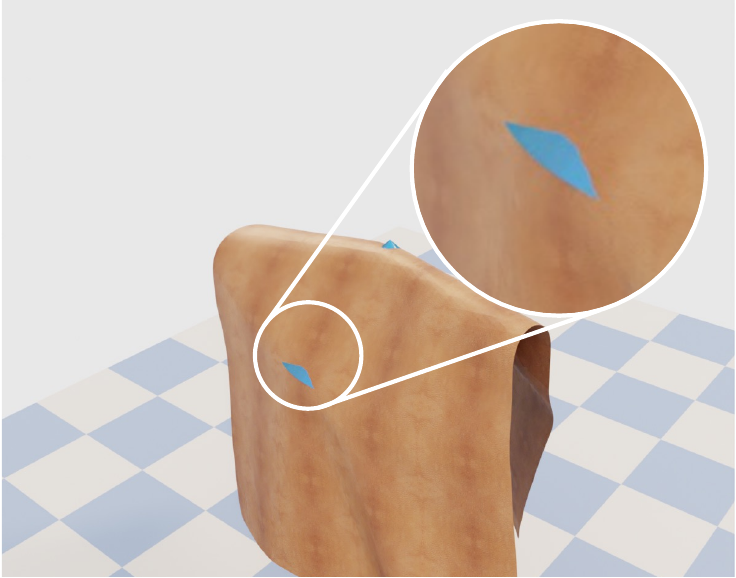}
        \caption{\textit{vertex-SDF\cite{fuhrmann2003vertex-sdf}}}
        \label{fig:sec4.2-vis-b}
    \end{subfigure}%
\hfill
    \begin{subfigure}{0.22\textwidth}
        \centering
        \includegraphics[width=1\textwidth]{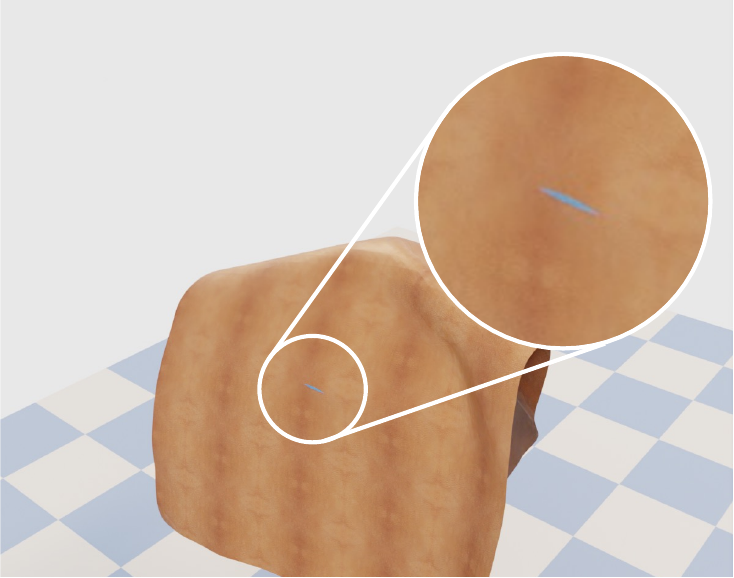}
        \caption{\textit{opt-SDF\cite{macklin2020optsdf}}}
        \label{fig:sec4.2-vis-c}
    \end{subfigure}%
\hfill
    \begin{subfigure}{0.22\textwidth}
        \centering
        \includegraphics[width=1\textwidth]{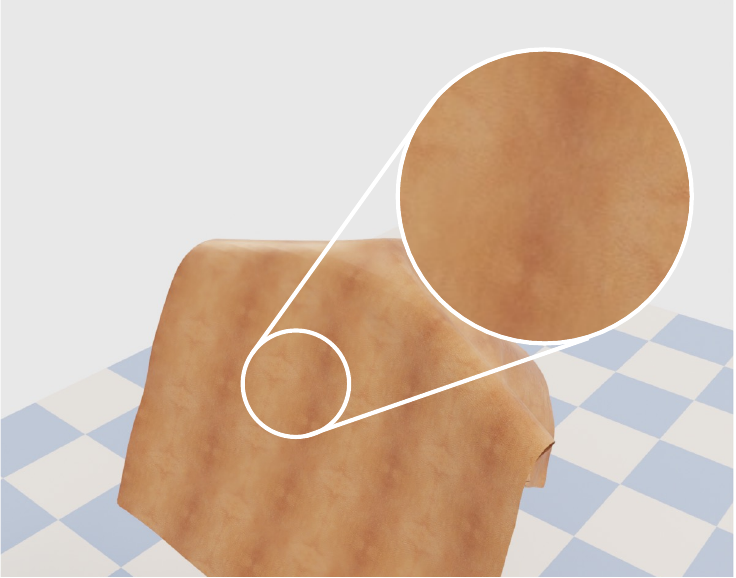}
        \caption{\textit{multi-res-SDF (ours)}}
        \label{fig:sec4.2-vis-d}
    \end{subfigure}%
\end{minipage}
\hfill
     \begin{subfigure}{0.5\textwidth}
        \centering
        \includegraphics[width=\textwidth]{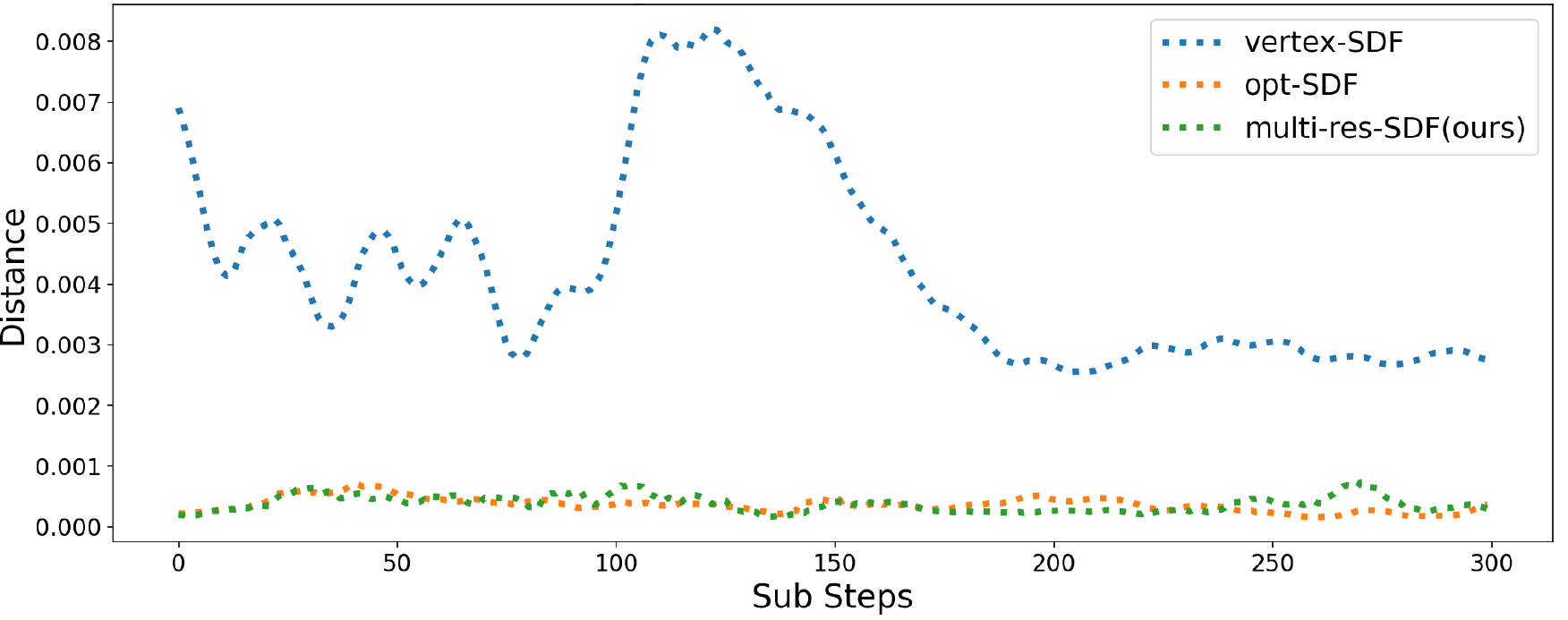}
        \caption{Average penetration distance}
        \label{fig:sec4.2-line-a}
    \end{subfigure}%
\hfill
      \begin{subfigure}{0.5\textwidth}
        \centering
        \includegraphics[width=\textwidth]{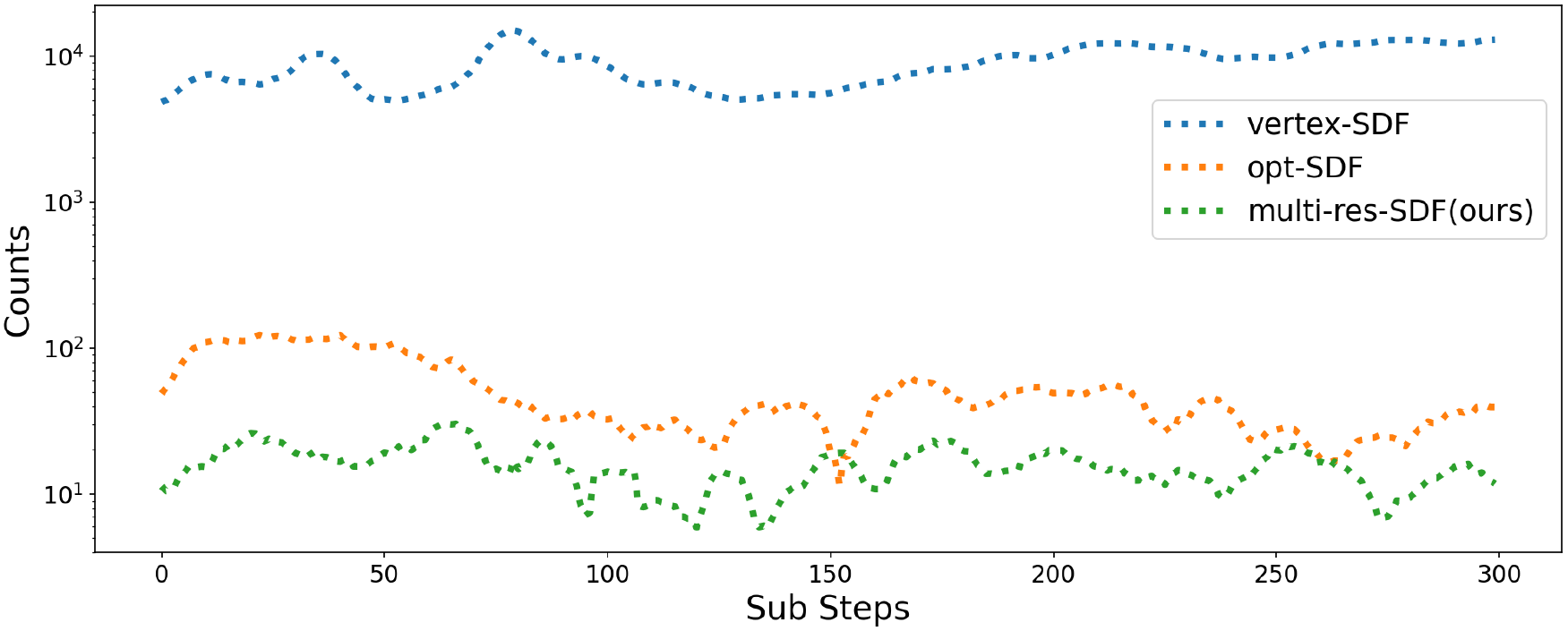}
        \caption{Penetration count}
        \label{fig:sec4.2-line-b}
    \end{subfigure}%

\caption{Qualitative comparison for contact handling. In (a), the skin initially lies above the cone and then falls due to gravity. (b) and (c) exhibit severe and slight penetration respectively, while (d) doesn't exhibit visible penetration.
Quantitative comparison for contact handling. In both metrics of (e) and (f), our method and \textit{opt-SDF}\cite{macklin2020optsdf} outperform \textit{vertex-SDF}\cite{fuhrmann2003vertex-sdf}. Top: \textit{multi-res-SDF} performs comparably to \textit{opt-SDF}. Bottom: \textit{multi-res-SDF} achieves nearly tenfold lower penetration count than \textit{opt-SDF}.}
\label{fig:sec4.2-contact-all}
\end{figure*}

\subsection{Contact}
\label{sec4.2:exp-contact}
\noindent \textbf{Baselines.}
We compare our method with other SDF-based strategies.
Fuhrmann \etal\cite{fuhrmann2003vertex-sdf} query the three vertices and three centers of its edges to determine the contact state, referred to as \textit{vertex-SDF}.
Macklin \etal\cite{macklin2020optsdf} perform an optimization on this triangle to find the local optimal point, abbreviated as \textit{opt-SDF}.
The former is widely employed\cite{verschoor2018softVR18,wang2024deepsimho,sorli2021fineISMAR}, while the latter is State Of The Art (SOTA).

\noindent \textbf{Simulation Setup.}
We reproduce a typical simulation scenario carried out in \cite{macklin2020optsdf}, including a cloth-like triangular mesh and a cone represented as the SDF\cite{sdfshader}.
The triangular mesh with 2,400 vertices serves as the skin and $\boldsymbol{C}^\mathrm{edge}$ and $\boldsymbol{C}^\mathrm{smo}$ are enabled.
The skin will fall and collide with this cone.
Macklin \etal\cite{macklin2020optsdf} set the number of iterations to 32 which means they query 32 points, so we also set the total number of query points to 32 within 2 sampling times for our multi-resolution method (abbreviated as \textit{multi-res-SDF}).
Within XPBD\cite{macklin2016xpbd} framework, we set the time step to 0.005 seconds and simulate a total of 30 steps, each of which is divided into 10 sub-steps.

\noindent \textbf{Evaluation Metrics.}
At the end of each sub-step, we uniformly sample 20 points on each triangle, count the number of points with negative signed distance values as the \textit{penetration count}, and calculate their \textit{average penetration distance} (without the sign).

\noindent \textbf{Results.} 
The qualitative results are shown in \cref{fig:sec4.2-contact-all}(a)-(d). 
There are severe penetrations in \textit{vertex-SDF}\cite{fuhrmann2003vertex-sdf} and slight in \textit{opt-SDF}\cite{macklin2020optsdf}, while our method does not have visible penetration.
For quantitative comparison, we compare the average penetration distance and penetration count of each sub-step in \cref{fig:sec4.2-contact-all}(e) and (f).
There is no doubt that \textit{multi-res-SDF} and \textit{opt-SDF} far surpass \textit{vertex-SDF}.
\cref{fig:sec4.2-line-a} shows that our method performs on par with \textit{opt-SDF}.
However, for penetration count (see \cref{fig:sec4.2-line-b}), we actually have nearly ten times fewer than \textit{opt-SDF}.
It is exactly because \textit{opt-SDF} has a higher number of penetrations that it reduces the average penetration distance.
In our approach, since each iteration of querying is performed by global sampling on the corresponding triangle, our approach can approximate the global optima.
This experiment simulates the collision between the skin and a sharp object, which is sufficient to demonstrate our remarkable ability in contact handling. 
Although it is challenging to mathematically prove that our method tends to achieve the globally optimal \textit{decision point}, the theoretical analysis mentioned in \cref{sec3.3:con} is consistent with current experimental results.
In addition, we further showcase the superior performance of \textit{multi-res-SDF} in HOI in \cref{sec4.3:exp-interaction}.

\begin{table*}[tb]
\centering
\caption{Quantitative comparison of contact methods in HOI experiments. 
We implement different collision detection methods including \textit{vertex-SDF}\cite{fuhrmann2003vertex-sdf}, \textit{opt-SDF}\cite{macklin2020optsdf}, and \textit{multi-res-SDF} (ours) in PhysHand and compare them together with ContactGen\cite{liu2023contactgen}. 
This comprises seven grasps involving six common objects.
Across all metrics of contacts, our method significantly surpasses other approaches with values approaching zero, demonstrating remarkable accuracy. PhysHand can run in real-time with GPU parallelization.
}

\label{tab:sec432}
\begin{tabular}{c|c|cccc}
\toprule
\multicolumn{2}{c}{\multirow{2}{*}{Object}}   & \multirow{2}{*}{\textit{ContactGen}\cite{liu2023contactgen}} & \multicolumn{3}{c}{PhysHand}\\
\cline{4-6}
\multicolumn{2}{c}{} & ~ & \multicolumn{1}{c}{\textit{vertex-SDF}\cite{fuhrmann2003vertex-sdf}}
& \multicolumn{1}{c}{\textit{opt-SDF}\cite{macklin2020optsdf}} & \multicolumn{1}{c}{\textit{multi-res-SDF} (ours)} \\ \hline
\multicolumn{1}{c|}{\multirow{3}{*}{glass-1}}    & \textit{count}$\downarrow$    & 143105              & 46285               & 156              & \textbf{3}                   \\
\multicolumn{1}{c|}{}                            & \textit{max} ($\mu m$)$\downarrow$  & 2503.9162           & 459.1479            & 6.2744           & \textbf{0.2956}              \\
\multicolumn{1}{c|}{}                            & \textit{avg.} ($\mu m$)$\downarrow$ & 1305.9646           & 114.9909            & 1.9033           & \textbf{0.2289}              \\ 
\multicolumn{1}{c|}{} & CPU/GPU \textit{time (ms)$\downarrow$} & - & \textbf{22}/- & 246/- & 2,517/\textbf{27} \\ \hline
\multicolumn{1}{c|}{\multirow{3}{*}{glass-2}}    & \textit{count}$\downarrow$    & 119067              & 63013               & 158              & \textbf{1}                   \\
\multicolumn{1}{c|}{}                            & \textit{max} ($\mu m$)$\downarrow$  & 2503.8602           & 442.6264            & 3.5655           & \textbf{0.0500}              \\
\multicolumn{1}{c|}{}                            & \textit{avg.} ($\mu m$)$\downarrow$ & 1150.6618           & 206.9566            & 1.9978           & \textbf{0.0500}              \\ 
\multicolumn{1}{c|}{} & CPU/GPU \textit{time (ms)$\downarrow$} & - & \textbf{21}/- & 243/- & 2,502/\textbf{27} \\ \hline
\multicolumn{1}{c|}{\multirow{3}{*}{toothpaste}} & \textit{count}$\downarrow$    & 71033               & 27435               & 30               & \textbf{1}                   \\
\multicolumn{1}{c|}{}                            & \textit{max} ($\mu m$)$\downarrow$  & 3106.4365           & 2313.6602           & 7.4098           & \textbf{0.0139}              \\
\multicolumn{1}{c|}{}                            & \textit{avg.} ($\mu m$)$\downarrow$ & 1068.7383           & 270.8561            & 1.7885           & \textbf{0.0139}              \\ 
\multicolumn{1}{c|}{} & CPU/GPU \textit{time (ms) $\downarrow$} & - & \textbf{21}/- & 169/- & 2,462/\textbf{26} \\ \hline
\multicolumn{1}{c|}{\multirow{3}{*}{phone}}      & \textit{count}$\downarrow$    & 287730              & 164933              & 2                & \textbf{1}                   \\
\multicolumn{1}{c|}{}                            & \textit{max} ($\mu m$)$\downarrow$  & 3323.5322           & 1916.7412           & 1.3571           & \textbf{0.0103}              \\
\multicolumn{1}{c|}{}                            & \textit{avg.} ($\mu m$)$\downarrow$ & 1411.4840           & 70.8230             & 0.7542           & \textbf{0.0103}              \\ 
\multicolumn{1}{c|}{} & CPU/GPU \textit{time (ms) $\downarrow$} & - & \textbf{15}/- & 227/- & 1,095/\textbf{12} \\ \hline
\multicolumn{1}{c|}{\multirow{3}{*}{apple}}      & \textit{count}$\downarrow$    & 79827               & 37927               & 268              & \textbf{0}                   \\
\multicolumn{1}{c|}{}                            & \textit{max} ($\mu m$)$\downarrow$  & 3294.8354           & 189.9071            & 3.7722           & \textbf{0}                   \\
\multicolumn{1}{c|}{}                            & \textit{avg.} ($\mu m$)$\downarrow$ & 1219.7463           & 78.8565             & 1.0504           & \textbf{0}                   \\ 
\multicolumn{1}{c|}{} & CPU/GPU \textit{time (ms) $\downarrow$} & - & \textbf{21}/- & 220/- & 1,995/\textbf{24} \\ \hline
\multicolumn{1}{c|}{\multirow{3}{*}{bowl}}       & \textit{count}$\downarrow$    & 86841               & 32120               & 190              & \textbf{11}                  \\
\multicolumn{1}{c|}{}                            & \textit{max} ($\mu m$)$\downarrow$  & 5718.0323           & 3129.6544           & 260.9675         & \textbf{0.0427}              \\
\multicolumn{1}{c|}{}                            & \textit{avg.} ($\mu m$)$\downarrow$ & 1788.3155           & 845.0303            & 35.4007          & \textbf{0.0093}              \\ 
\multicolumn{1}{c|}{} & CPU/GPU \textit{time (ms) $\downarrow$} & - & \textbf{18}/- & 145/- & 1,480/\textbf{16} \\ \hline
\multicolumn{1}{c|}{\multirow{3}{*}{box}}        & \textit{count}$\downarrow$    & 163436              & 58827               & 820              & \textbf{7}                   \\
\multicolumn{1}{c|}{}                            & \textit{max} ($\mu m$)$\downarrow$  & 5217.0626           & 989.9247            & 0.4664           & \textbf{0.0050}              \\
\multicolumn{1}{c|}{}                            & \textit{avg.} ($\mu m$)$\downarrow$ & 2331.5708           & 133.4961            & 0.0110           & \textbf{0.0016}              \\ 
\multicolumn{1}{c|}{} & CPU/GPU \textit{time (ms) $\downarrow$} & - & \textbf{17}/- & 201/- & 1,073/\textbf{12} \\
\noalign{\global\setlength{\arrayrulewidth}{0.75pt}}\cline{1-6} 
\noalign{\global\setlength{\arrayrulewidth}{0.4pt}} 
\end{tabular}
\end{table*}

\subsection{Interaction}
\label{sec4.3:exp-interaction}
Although a series of works on hand simulation have been proposed\cite{ott2010twohand,jacobs2012god,jacobs2011softVR11,hirota2016interactionVR16,verschoor2018softVR18}, they are either inaccessible or hard to use, leading to unavailable of comparison.
The previous experiments have illustrated the capabilities of physics constraints and contact handling. 
Next, we will show the significant value of the full model on interaction.
We observe that a recent spate of studies on grasp synthesis has emerged\cite{karunratanakul2021halo,liu2023contactgen,christen2022d-grasp}, which considerably enriches the datasets for HOI.
Despite achieving remarkable progress in grasp pose generation, these approaches often fall short due to an inherent limitation: their inability to handle contact deformations. Consequently, the synthesized grasps always exhibit penetration. 
This artifact diminishes the realism of HOI, which is exactly what PhysHand has addressed.
Thus, the first experiment is about improving the grasp data.
Subsequently, we implement the same experiment as the first one with \textit{vertex-SDF}\cite{fuhrmann2003vertex-sdf} and \textit{opt-SDF}\cite{macklin2020optsdf}, as introduced in \cref{sec4.2:exp-contact}.
This evaluation aims to underscore PhysHand's enhancements to managing physical interactions with higher accuracy.

\begin{figure*}[htbp]
\centering
    \begin{subfigure}{\textwidth}
        \centering
        \includegraphics[width=\textwidth]{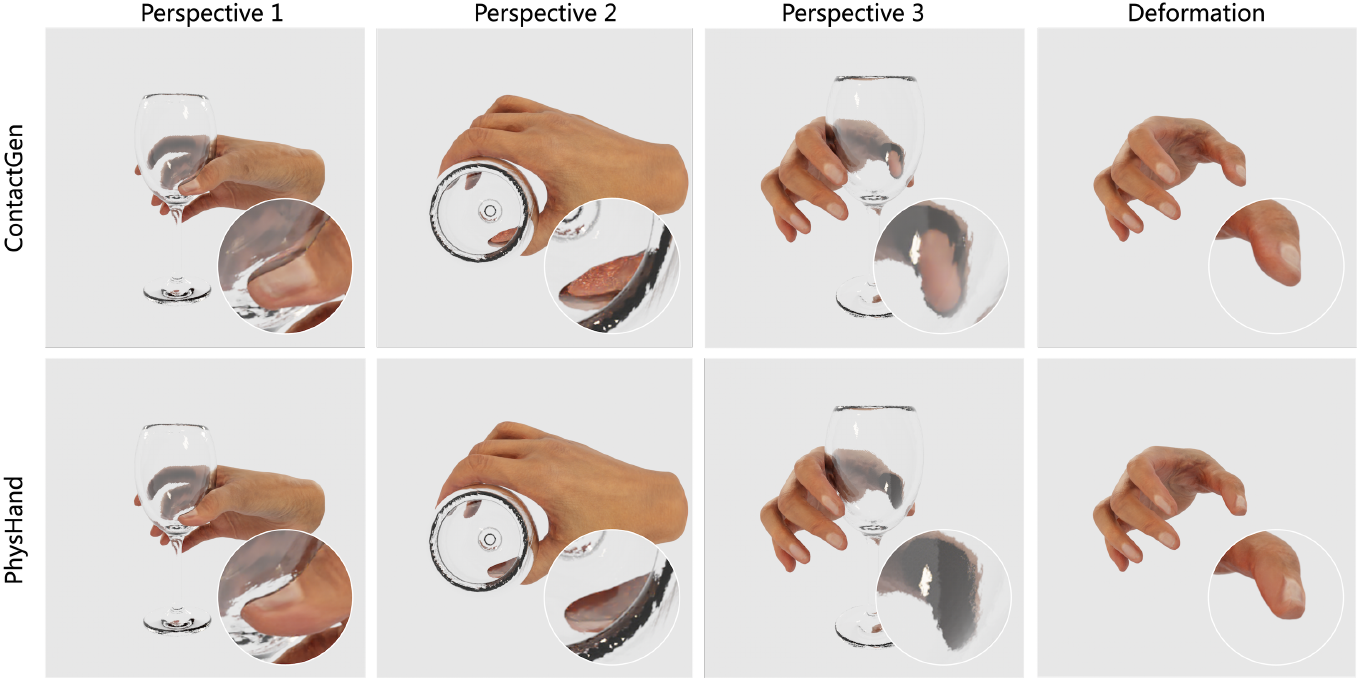}
        \caption{\textit{glass-1}}
        \label{fig:sec4.3-exp1-bottle1}
    \end{subfigure}%
    \hfill

     \begin{subfigure}{\textwidth}
        \centering
        \includegraphics[width=\textwidth]{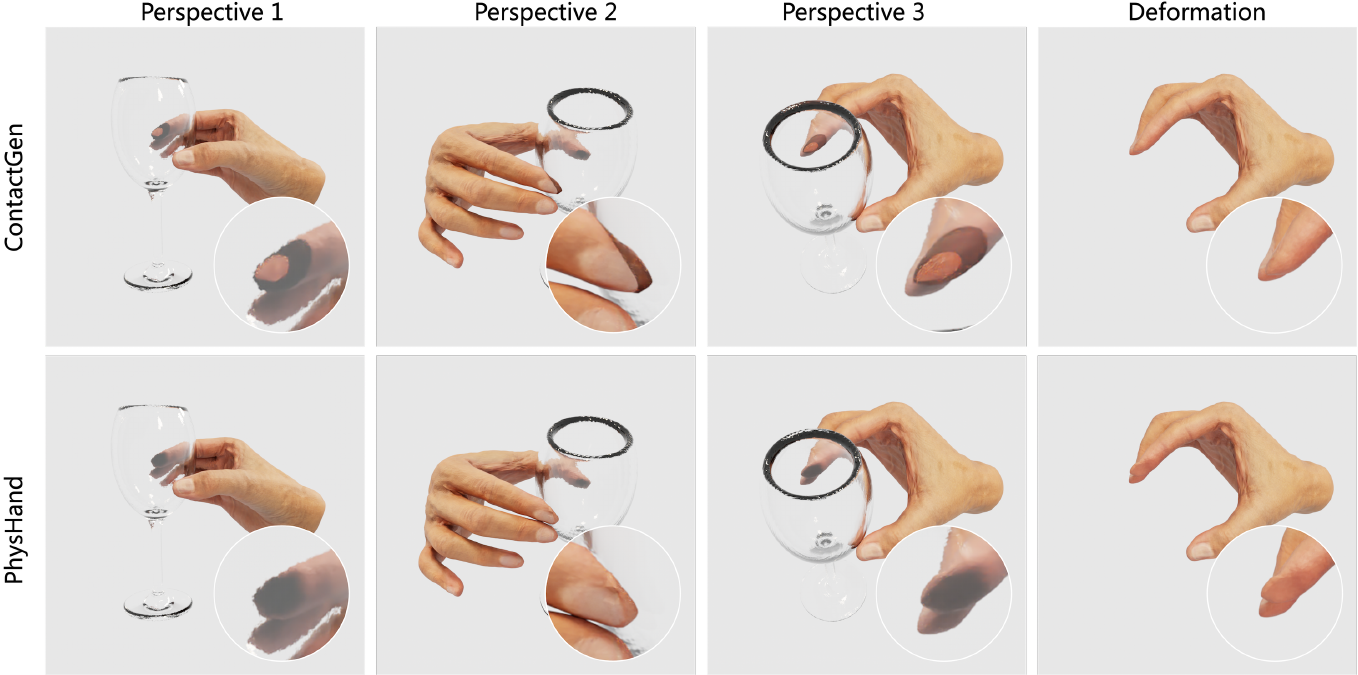}
        \caption{\textit{glass-2}}
        \label{fig:sec4.3-exp1-bottle2}
    \end{subfigure}%
    
    \caption{Qualitative comparisons to \textit{ContactGen}\cite{liu2023contactgen}.
    (a) and (b) depict two different grasps of the same glass. 
    Columns one through three represent three perspectives, while the fourth column illustrates the pure hand in the third column's view.
    As a result, PhysHand achieves realistic deformations, while \textit{ContactGen} suffers from distortions due to penetrations.
    This experiment demonstrates the potential application of our PhysHand in enhancing the quality of grasps datasets.
    }
\label{fig:sec4.3-exp1}
\end{figure*}

\noindent \textbf{Qualitative Comparison of Improving Grasp Data.}
\label{sec4.3.1}
We replicate \textit{ContactGen}\cite{liu2023contactgen}, a SOTA model in generating hand-object grasps.
Although it proposes a novel contact feature representation of the contact map, hand part, and direction, the generated grasp suffers from penetration issues.
Therefore, we make comparisons to \textit{ContactGen} on two typical grasp samples to demonstrate the significance of PhysHand.
Given an object, we first fit its SDF using a combined analytic function similar to \cite{sdfshader}.
Subsequently, \textit{ContactGen} generates the grasping pose for this object, serving as the input pose for PhysHand. 
Finally, we conduct the simulation.
For each grasp, we present detailed views from three perspectives along with a single rendering depicting the scene only with the deformed hand, as shown in \cref{fig:sec4.3-exp1}.
Despite \textit{ContactGen} creating two excellent grasps, there are obvious penetrations.
PhysHand demonstrates its unique value in this scenario by providing more authentic deformations and overcoming visible penetrations.
Similar situations are also prevalent in robotics\cite{urain2023se3,khargonkar2023neuralgrasps,li2023rdf}.
They concentrate on pose while neglecting the realism of deformation, even considering penetration as a criterion for successful grasping.
Nevertheless, this should not be attributed to these works, as the underlying reason lies in the absence of suitable simulation models, which is exactly what PhysHand accomplishes.

\begin{figure*}[htbp]
\centering
     \begin{subfigure}{\textwidth}
        \centering
        \includegraphics[width=\textwidth]{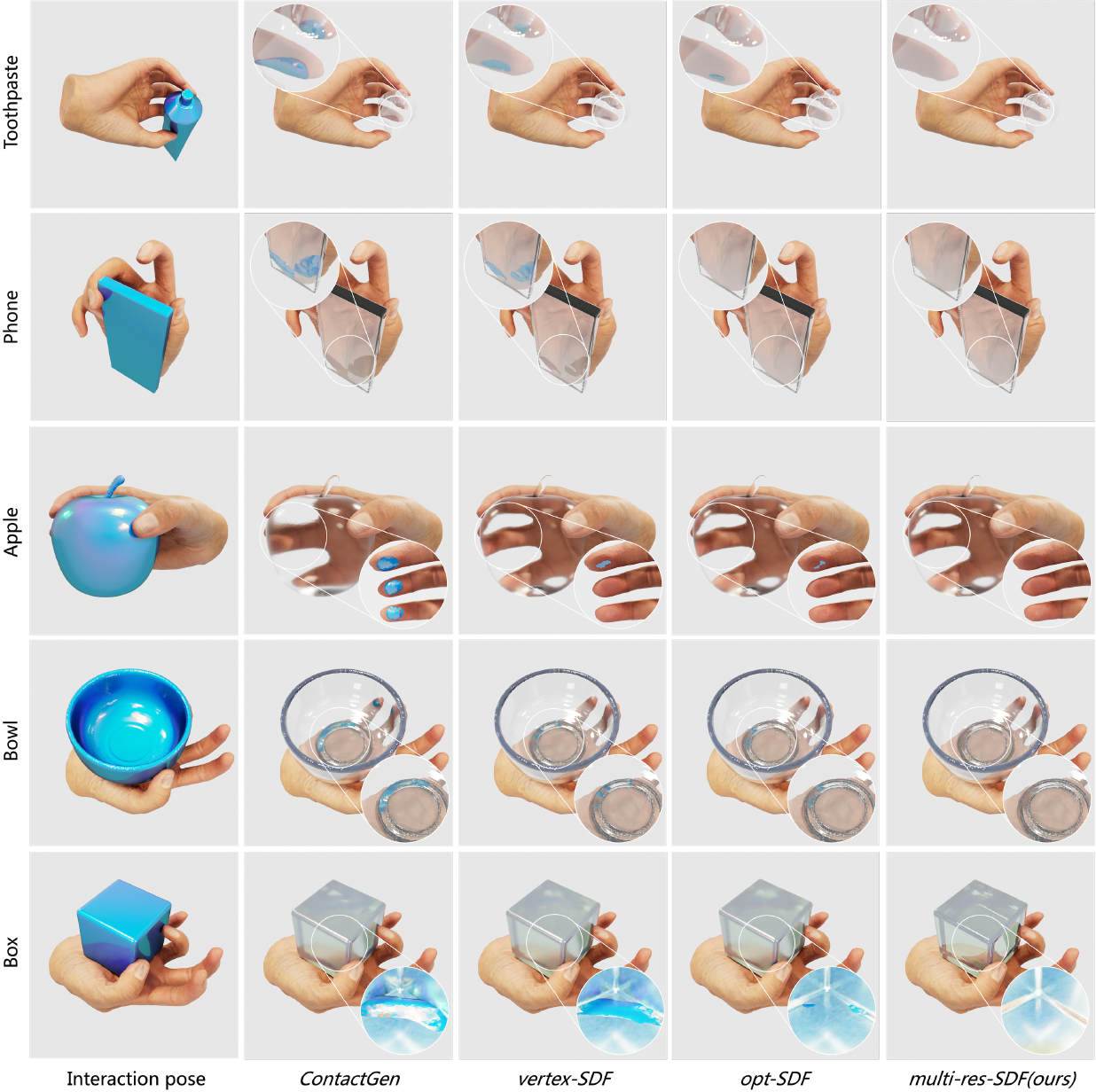}
        \label{fig:sec4.3-exp2-toothpaste}
    \end{subfigure}%
\caption{Qualitative results of \textit{toothpaste},\textit{phone},\textit{apple},\textit{bowel} and \textit{box}.
The first and second columns are from \textit{ContactGen}\cite{liu2023contactgen}, while the others are from PhysHand with different contact handling methods.
For better visualization, the results are rendered transparently, which is unrelated to the material. 
Specifically, the dark blue regions indicate penetrations in all graphs except for the first column.
PhysHand with \textit{multi-res-SDF} exhibits excellent performance with no visual penetration across all examples.    
}
\label{fig:sec4.3-exp2}
\end{figure*}

\noindent \textbf{Further Comparison on Contact Handling.}
\label{sec4.3.2}
Apart from \textit{glass-1} and \textit{glass-2} (\cref{fig:sec4.3-exp1}), another five examples are carried out, employing the contact handling strategies mentioned in \cref{sec4.2:exp-contact}, namely \textit{vertex-SDF}\cite{fuhrmann2003vertex-sdf}, \textit{opt-SDF}\cite{macklin2020optsdf}, and \textit{multi-res-SDF} (ours).
The quantitative result is illustrated in \cref{tab:sec432}, while \cref{fig:sec4.3-exp2} is the corresponding visualization.
The evaluation metrics concerning the vertices of the object include the count of penetration, the maximum penetration depth, and the average penetration depth, abbreviated as \textit{count}, \textit{max}, and \textit{avg.}, respectively.
A lower value of the evaluation metric signifies superior performance.
We simulate a total of 36 steps to ensure convergence, with each step consisting of 10 sub-steps. 
For these metrics, we calculate the values of each sub-step within the last time step.
Regarding measurements, we assume that the length of the hand is 180 millimeters, which is consistent with reality. 
Since the metric results of our method are extremely low, the units for \textit{avg.} and \textit{max} are in micrometers.
As a reference, we additionally compare with \textit{ContactGen}\cite{liu2023contactgen}. 
In \cref{tab:sec432}, our approach exhibits remarkable performance, as evidenced by its superiority over other methods, with values approaching zero across all metrics. 
Such a low level of penetration is sufficient enough to support tasks that involve detailed contact, \textit{e.g.,} precise manipulation of screws, knobs, \textit{etc.}
In \cref{fig:sec4.3-exp2}, the dark blue region in the other images represents penetration except for the first column.
Compared to \textit{ContactGen}, PhysHand can eliminate penetration and generate realistic deformations.
As for contact handling methods, PhysHand with \textit{multi-res-SDF} is the most accurate, with no visible penetration in all examples.

\noindent 
\textbf{Run-time Performance.}
We test the run-time performance on the 12th-Gen-Intel-i7-12700/RTX-3080Ti-10GB.
The metric of CPU/GPU \textit{time} in \cref{tab:sec432} means the contact handling time per frame.
We omit the parallel testing of \textit{vertex-SDF} due to severe penetration. 
\textit{opt-SDF} does not support parallelization due to the sequential querying.
Querying in \textit{multi-res-SDF} (ours) is independent of each point, allowing for parallel implementation on GPU.
Though our speed is slowest on a single CPU due to uniform sampling and SDF value sorting, our parallel deployment greatly outperforms others, achieving high real-time performance.
Beyond \cref{tab:sec432}, for \textit{multi-res-SDF}, the total simulation time per frame is 55ms, 55ms, 61ms, 38ms, 59ms, 37ms, and 48ms. Therefore, our work is qualified for real-time application.

\begin{figure*}[t]
\centering
    \begin{subfigure}{1.0\textwidth}
        \centering
        \includegraphics[width=\textwidth]{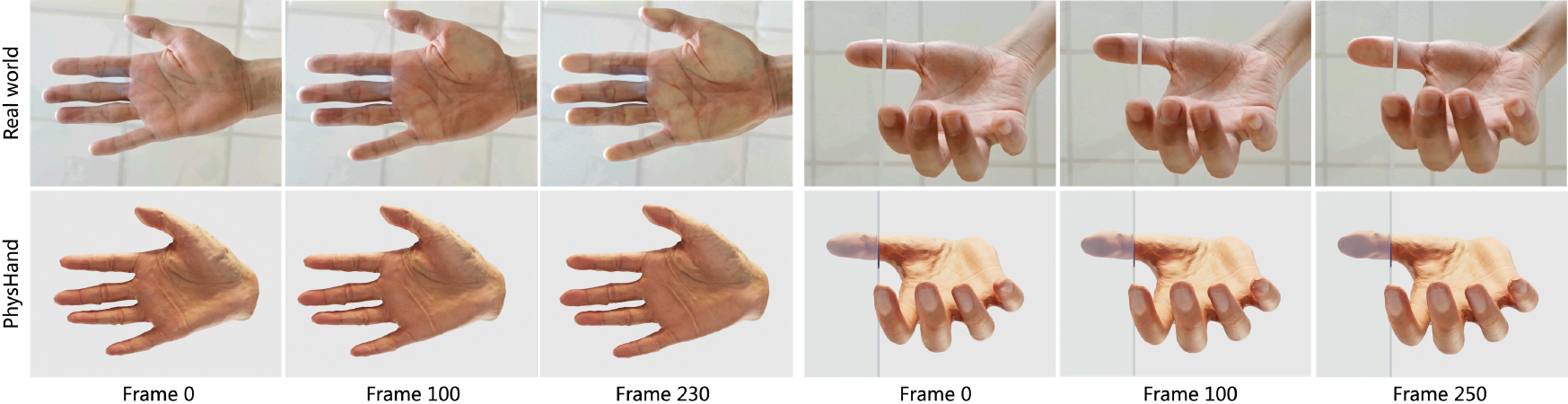}
    \end{subfigure}%
    \caption{
    This experiment clearly demonstrates that our PhysHand is able to achieve deformations remarkably close to those observed in the real world.
    We render the result without the plate on the left example for better visualization.
    }
\label{fig:sec4.4-real}
\end{figure*}

\subsection{Compared to the deformation in the real world}
\label{sec4.4:real}
It is hard to measure real-world deformations, so we devised a simple setup to assess the authenticity of our simulation outcomes visually.
This setup consists of a square, transparent plate laid horizontally. 
A hand presses down vertically from above the plate while a camera records this process from beneath.
In PhysHand, we replicate this procedure by manually setting the hand's pose and the rigid motion of the root joint.
\cref{fig:sec4.4-real} showcases two pressing examples. 
It is quite similar between the real world and our simulation in the shape and distribution of deformation regions.

\section{Discussion}
\noindent \textbf{Limitations.}
\label{sec5.1}
First, since self-collision is not considered in this work, self-penetration between triangles might occur after solving the collision constraint. However, the other constraints will also generate displacements, thus occasional self-penetration does not impact the convergence of the simulation.
Second, PhysHand primarily focuses on deformation rather than actuation, thus we treat muscles and fat as the same soft tissue, which hinders further enhancement in realism.
Finally, the absence of friction prevents PhysHand from actively grasping objects. 
However, this can be easily addressed by integrating friction models\cite{harmon2009friction-1,kaufman2008friction-2}. 

\noindent \textbf{Future Works.}
\label{sec5.2}
We aim to develop PhysHand as a plugin compatible with popular graphics engines such as Unreal or Unity to share our work.
Moreover, we plan to incorporate muscles into PhysHand, for both deformation and actuation. 
Currently, Physhand supports the unidirectional deformation of the hand model. It is valuable to consider the bidirectional interaction that applies force to the objects.

\section{Conclusion}
\noindent We present a novel hand simulation model, PhysHand, capable of faithfully reproducing the deformations in HOI.
Our approach introduces three key innovations to ensure authenticity.
First, our method considers a layered geometry that fully encompasses the skeleton, flesh, and skin.
Second, we employ a flexible and scalable dynamics framework that enables the design of 
physics constraints for skin and flesh to ensure their distinct non-rigid deformations.
Lastly, a multi-resolution querying strategy is proposed to enhance SDF-based contact handling.
We hope PhysHand can make HOI more authentic.

\section*{ACKNOWLEDGMENT}
This work is Sponsored by CIE-Tencent Robotics X Rhino-Bird Focused Research Program.

\bibliographystyle{eg-alpha-doi} 
\bibliography{egbibsample}

\end{document}